\documentclass[manuscript,screen]{acmart}
\AtBeginDocument{%
  }

\setcopyright{acmlicensed}
\copyrightyear{2018}
\acmYear{2018}
\acmDOI{XXXXXXX.XXXXXXX}
\acmConference[Conference acronym 'XX]{Make sure to enter the correct
  conference title from your rights confirmation email}{June 03--05,
  2018}{Woodstock, NY}
\acmISBN{978-1-4503-XXXX-X/2018/06}

\usepackage{tabularx}
\usepackage{xcolor}
\begin{document}

\title{Losing One’s Story: How Vulnerable Users Experience Harm in Online Support Seeking}



\author{Yingfan Zhou}
\affiliation{%
  \institution{The Pennsylvania State University}
  \country{USA}
}
\author{Priya Kumar}
\affiliation{%
  \institution{The Pennsylvania State University}
  \country{USA}
}
\author{Anna Squicciarini}
\affiliation{%
  \institution{The Pennsylvania State University}
  \country{USA}
}

\author{Sarah Rajtmajer}
\affiliation{%
  \institution{The Pennsylvania State University}
  \country{USA}
}
\begin{abstract}
Online support communities are a critical resource for individuals facing distress, stigma, and limited offline support. However, these spaces are not uniformly supportive, particularly for users with little margin for error. In this paper, we examine how vulnerable users experience harm within online support-seeking interactions. Through 25 semi-structured interviews with Reddit users, we show that support seeking is often a constrained practice shaped by structural vulnerability. We introduce the concept of \emph{narrative harm} to describe how participants experience 
  harm through losses of narrative authority, coherence, and space. Their personal disclosures are questioned, reframed, or displaced by others, reflecting asymmetries in voice, credibility, and interpretive power embedded in platform dynamics and moderation regimes. In response, users engage in defensive strategies, including selective self-disclosure, self-silencing, identity separation, and informal mutual aid, shifting the burden of safety onto those already most vulnerable. Our findings highlight how current platform designs insufficiently account for situated vulnerability and redistribute discursive power away from support seekers. 
  We discuss implications for the design of online support systems that better preserve narrative integrity and reduce the burden of safety labor.  
\end{abstract}



\maketitle
\section{Introduction}
Online communities have become a key resource for individuals seeking support related to a wide range of personal issues, e.g., health \cite{skeels2010catalyzing, de2014seeking, yang2019channel,young2019girl}, employment \cite{tomprou2019career}, and trauma \cite{andalibi2019happens,andalibi2016understanding}. For individuals with limited offline resources, these spaces are often  a strong and needed resource for support and information sharing.  
The broad reach of social networks diversifies information sources and increases the likelihood of accessing higher-quality support, including responses from individuals with relevant expertise \cite{wright2003health,kramer2021strength,burke2013using}. 
Ample prior work suggests that online support can mitigate distress, boost well-being, and improve physical and mental health \cite{jourard1971self,derlega1993self,house1988structures, thoits1995stress}.

However, online support is not uniformly beneficial.
Prior research has documented more overt forms of harm that occur within support-seeking environments \cite{yin2025understanding}. Users do not always receive the support they seek, and failed attempts to obtain support can intensify their distress \cite{de2014mental, forest2012social}. Some online support groups foster negative emotional contagion or social comparison \cite{barta2023similar}, while others may exclude help-seekers due to skepticism about their identity or the authenticity of their disclosures \cite{wakefield2013meta}. 

Such harm is not evenly distributed \cite{burger2025saying,agha2024tricky} and online spaces are not equally substitutable across users \cite{10.1145/3375195}. For individuals with limited offline resources who lack accessible support, information, or institutional assistance offline, online communities may not function merely as supplementary spaces, but as crucial tools  to  seek help that is difficult to access elsewhere \cite{blank2018benefits}. We draw on the concept of social vulnerability from interdisciplinary work in sociology \cite{morrow1999identifying,fothergill2004poverty}, anthropology \cite{oliver1996anthropological}, and environmental science \cite{cutter2012vulnerability,cutter2012social}, where it is used to describe differential exposure to hazards as well as unequal capacities to cope with, recover from, or prevent harm. In this study, we focus on socially vulnerable individuals to understand how resource constraints shape users’ reliance on online support and their capacity to cope with, recover from, or prevent online harm. 

Harm is understood as a setback to a person’s fundamental interests, well-being, or rights \cite{feinberg1989moral}. Prior work has documented harms that produce physical, emotional, relational, and financial loss \cite{andalibi2021sensemaking,karizat2025laboring, simpson2021you}. Building on this foundation, we study more subtle forms of harm that emerge within everyday online support interactions.  Rather than treating offline vulnerability as directly producing online harm, we ask how social vulnerability shapes users’ dependence on online support, their exposure to harmful interactions  in the course of seeking that support, and how these experiences in turn affect their subsequent support-seeking behavior. Our research is structured around three core research questions (RQs):
\begin{enumerate}
    \item[\textbf{RQ1:}] How do vulnerable users experience harm in everyday online support interactions?
    \item[\textbf{RQ2:}] How do self-disclosure and support-seeking practices shape these experiences?
    \item[\textbf{RQ3:}] How do these experiences influence subsequent support-seeking behaviors?
\end{enumerate}

We address these questions through 25 semi-structured interviews with Reddit users. As individuals disclose personal experiences to seek help, they become exposed to questioning, reinterpretation, and misalignment, sometimes losing control over how their stories are understood. We conceptualize this as \emph{narrative harm}: the loss of narrative authority, coherence, and space in support-seeking interactions. Rather than stemming from overt toxicity, such harm often arises through subtle shifts in interpretation, credibility, and attention.

Our contributions are threefold: (1) we show how offline scarcity and stigma constrain support seeking and heighten vulnerability to narrative harm; (2) we unpack how such harm unfolds through reinterpretation and credibility shifts within interactions; and (3) we trace how users adapt, by adopting strategies such as self-silencing, selective disclosure, identity separation, and informal mutual aid. Taken together, our findings reveal sociotechnical gaps in current platform design and suggest new directions to better preserve and account for situated vulnerability.

\section{Related Work}

\subsection{Online Harm  and Vulnerability}

Following prior work on online harm, we use the term online harm broadly to encompass a range of users' experiences \cite{agrafiotis2018taxonomy,scheuerman2021framework,livingstone2013online}. Existing research has generally classified harmful behaviors based on the nature and extent of the harm caused by individuals’ online actions or the information they share \cite{agrafiotis2018taxonomy,scheuerman2021framework}, e.g., 
physical, emotional, relational, and financial harms \cite{scheuerman2021framework}. 
A substantial literature has focused on toxic interactions, including harassment \cite{musgrave2022experiences,rubin2020fragile,karusala2017women}, cyberbullying \cite{singh2017they, maqsood2021they,iivari2021chi}, hate speech \cite{thomas2022s,phadke2020many}, and flaming \cite{jane2015flaming, take2024stoking} on social media platforms. Another line of research has investigated problematic online content that may not involve direct interpersonal exchanges yet still causes harm or negative effects to readers or platforms, such as misinformation \cite{lazer2018science, gunasekara2025influence,konstantinou2025behavior} and information-induced trauma \cite{scott2023trauma, zheng2024s, ali2024m, doyle2024hate}.

Authors have proposed a range of interventions to mitigate online harm arising from user-generated content on social media, which takes the form of text, images, or videos \cite{whittaker2015cyberbullying}. Prior work has proposed both proactive and reactive interventions, including UI nudge designs \cite{masrani2023slowing,gencc2024situating,agha2023strike}, 
content moderation systems \cite{cai2024content,scott2023trauma,zhang2025fewer}, 
reporting mechanisms and blocking functions 
\cite{wei2023there,goyal2022you,baumler2025towards}. Software tools have been developed to empower users in the wake of harassment \cite{xiao2025snugglesense, kim2024respect}, while peer support networks and aid groups provide support to both victims \cite{doan2025design} and moderators \cite{uttarapong2024moderatorhub, mayworm2024online}. 

However, the subtle, subjective, and context-dependent nature of online harm continues to challenge existing mitigation efforts. Empirical studies show that definitions of online harm vary across users, communities, and cultural backgrounds \cite{kim2021you, haimson2021disproportionate, davani2024disentangling}. Likewise, perceptions of the same terms or definitions vary when applied to real-world examples \cite{hartmann2025lost, chang2025opaque,thebault2023diverse}. Furthermore, context-dependent harms often lack explicit policy violations, making them difficult to detect \cite{williams2019desensitization,ryu2025exploring, heung2025ignorance, yang2025m}. 

Finally, risks of encountering online harm are not evenly distributed across users. Certain groups, such as women \cite{burger2025saying, musgrave2022experiences,vitak2017identifying}, LGBTQ+ \cite{taylor2024cruising,tanni2024examining,burger2025saying}, racial minorities \cite{musgrave2022experiences,williams2019desensitization,williams2025can}, children \cite{agha2024tricky,hardwick2025they,oguine2025online}, and people with disabilities \cite{heung2025ignorance,lyu2024got,heung2024vulnerable}, experience harm more frequently and intensely than others. Critically, equality does not imply equity \cite{mouzannar2019fair}; interventions that treat all users identically fail to address the specific challenges faced by these populations \cite{heung2024vulnerable,10.1145/3701186,anuyah2023characterizing}. Inadequate mechanisms and scarce support resources may even cause further harm to these groups \cite{baumler2025towards}.

To understand these uneven experiences of online harm, we draw upon the concept of "vulnerability" in interdisciplinary research spanning sociology \cite{morrow1999identifying,fothergill2004poverty}, anthropology \cite{oliver1996anthropological}, and environmental science \cite{cutter2012vulnerability,cutter2012social}, which broadly refers to differences in exposure to harm and in the capacity to anticipate, cope with, recover from, or prevent adverse outcomes. To promote diversity and inclusion in design, HCI research has increasingly examined various vulnerable groups through taxonomic and population-specific approaches \cite{10.1145/3375195,heung2024vulnerable,anuyah2023characterizing}. However, \citet{10.1145/3710935} caution that population-based approaches can obscure within-group diversity, overlook the conditions that produce vulnerability, and reinforce othering. They point out that vulnerability is a dynamic condition shaped by individual circumstances, interpersonal relations, institutions, sociocultural structures, and temporal change.

Following this perspective, we treat vulnerability as a contextual condition rather than an inherent demographic attribute. Particularly, we locate vulnerability in situations where users' offline resources, support, or institutional access are constrained, shaping their reliance on online support spaces and their capacity to cope with online harm.

\subsection{Self-Disclosure and Narrative}

Self-disclosure refers to the voluntary sharing of personal information \cite{jourard1958some,cozby1973self}. On social media platforms, users actively select, organize, and construct their experiences to clearly present themselves to an audience and fulfill social objectives \cite{lee2023online,wingate2020influence,lee2013lonely}. These actions point toward the concept of narrative \cite{pasupathi2009tell}.

A narrative refers to a structured account of connected events that forms a meaningful story \cite{hinchman1997memory}. People make sense of their experiences by organizing events into narratives that establish context, connect actions over time, and convey meaning \cite{ricoeur1980narrative, ricoeur1991human}. This process of creating and interpreting stories forms a cornerstone of the human experience. Individuals construct a sense of self by organizing their experiences into a coherent life story through which they interpret what has happened and locate themselves within a broader framework of meaning \cite{mcadams2013narrative,mclean2015personal}. 

Critically, this process is not something individuals accomplish alone. The listeners actively participate in shaping how a story is told, interpreted, and understood \cite{pasupathi2009development}. The responses of the listeners feed back into how a narrator understands and re-presents their own experience \cite{pasupathi2015being,jennings2014intricate}. In this sense, the narrative construction of “who I am” is also social: it depends not only on the teller's capacity to tell but also on the audience's willingness to listen, recognize, and engage \cite{pasupathi2015being, pasupathi2009development}.

Taken together, online self-disclosure is not merely the sharing of personal information but a form of narrative construction through which individuals organize, interpret, and present their experiences, thereby constructing an online sense of self through interaction with listeners.

\subsection{Online Support Seeking}

Online communities have become an important space for individuals to seek social support across domains such as health \cite{skeels2010catalyzing, de2014seeking, yang2019channel,young2019girl}, mental well-being \cite{pretorius2020searching,sharma2018mental,de2014mental,progga2023understanding}, identity \cite{augustaitis2021online}, stigma \cite{rho2017class,andalibi2019happens}, and career development \cite{tomprou2019career}. Social support refers to the perception or experience of receiving care or assistance from others \cite{house1988structures}, and is commonly categorized as emotional, informational, or instrumental support \cite{rodriguez1998social}. Compared to offline settings, online environments lower barriers to accessing support by enabling public, persistent, and scalable interactions, allowing users to share experiences and resources more widely \cite{wright2003health,burke2013using,kramer2021strength, 10.1145/3633068}.

Effective support seeking in these contexts often depends on users’ willingness to disclose personal experiences. 
Expressing feelings, thoughts, and lived experiences provides critical context that enables others to interpret needs and offer relevant support \cite{lee2023online,wingate2020influence,lee2013lonely}. Prior work shows that open expression of negative experiences can reduce distress and contribute to improved mental health outcomes \cite{jourard1971self,derlega1993self,zhang2017stress,luo2020self}. As a result, online support communities tend to encourage higher levels of self-disclosure than general-purpose online spaces \cite{barak2007degree,lee2023online}.

At the same time, online self-disclosure is not solely an individual decision. The form and extent of disclosure are shaped by platform affordances and social context.  Users must continually balance how much to disclose, what support they hope to receive, and the risks they are willing to assume \cite{murphy2025sharing, sultana2022imagined, jo2024understanding}. These negotiations are influenced by community norms, social cues, and the behaviors modeled by others, which can encourage deeper or more emotionally charged disclosures than users might otherwise choose \cite{yang2019channel, pan2018you}. Features such as anonymity, communication channels, and prevailing community norms further shape how and how much users choose to reveal about themselves \cite{yang2019channel, pan2018you}.

Increased disclosure can heighten vulnerability \cite{karizat2025laboring, simpson2021you}. Responses that stigmatize or dismiss support-seekers, as well as excessive or inappropriate emotional reactions, can evoke shame and reduce individuals’ willingness to seek support in the future \cite{malik2010they,griffiths2011seeking}. Communities may unintentionally enable negative emotional contagion and foster harmful social comparison \cite{barta2023similar}. In addition, misinformation and harmful speech embedded within support exchanges can misguide users or cause psychological harm, particularly in vulnerable communities \cite{karusala2021courage,augustaitis2021online}. 

Together, these findings suggest that while self-disclosure is foundational to online social support, it also creates conditions under which harm may emerge. However, prior work has primarily conceptualized this risk in terms of exposure to harmful content or toxic interactions, leaving under-explored ways in which harm unfolds through the interpretation, reshaping, and contestation of experiences.

\subsection{Online Support Seeking on Reddit}

Prior research has shown that differences in platform design can shape strategies and interactions with other community members during online support seeking \cite{andalibi2019happens,de2014mental,andalibi2017sensitive}. 
Reddit is organized into topic-specific communities ("subreddits"), each governed by its own rules, norms, and moderation practices \cite{yang2019channel,rajadesingan2020quick}. This decentralized structure enables users to seek support across diverse contexts, while also producing variation in how support seeking is interpreted and responded to. Moderators play a role in shaping norms around disclosure, privacy, and appropriate interaction, though enforcement is often uneven \cite{malik2010they, saha2020understanding}.

Pseudonymity is a key feature of Reddit. Reddit was not originally designed with support seeking as a primary goal \cite{singer2014evolution}.  However, its semi-anonymous structure encourages self-disclosure, for example around stigmatized, sensitive, or distressing experiences \cite{leavitt_this_2015,ammari2019self}. As account creation is low-cost, users can employ “throwaway” accounts, or temporary accounts used briefly and then abandoned \cite{leavitt_this_2015,ammari2019self}. “Throwaway” accounts enable users to move across social contexts and adopt forms of expression that are difficult under stable, real-name identities. Reduced identifiability eases the tension between disclosure and privacy concerns \cite{de2014mental,andalibi2016understanding}. 
However, pseudonymity can also lead to disinhibition, potentially exposing support seekers to harmful interactions even as they reach out for help \cite{lapidot2012effects,andalibi2021sensemaking,karizat2025laboring, simpson2021you}. Furthermore, pseudonymity can blur boundaries around expertise related to the advice being offered \cite{pan2020say,yang2019channel}.

Reddit's voting system (upvotes and downvotes) together with an account-level reputation metric (“karma”) are designed to facilitate community-based content curation and reputation signaling. Highly upvoted posts become more visible in subreddit feeds, while downvoted content becomes less visible. Karma accumulates over time and serves as a proxy for user credibility and standing. In support-seeking contexts, these mechanisms shape both the visibility of disclosed experiences and the perceived legitimacy of responses. Higher karma can lend credibility to otherwise anonymous contributors \cite{yin2025understanding,de2014mental}. 

Collectively, these features make Reddit a valuable context for exploring how individuals seek and provide support, as well as how they navigate potential risks of harm within online support environments.

\section{Methods}
We conducted 25 semi-structured interviews with Reddit users between June and August 2023. Reddit provides a suitable setting for studying our research questions due to its community-oriented structure, topic-specificity, and anonymity affordances. These features enable users to seek and exchange support across diverse contexts \cite{de2014mental,yin2025understanding}. 


\subsection{Participant Recruitment}
To recruit participants, we first identified subreddits focused on providing support. We began with one of the largest support communities on the platform, which maintains a resource page of other support-focused subreddits. We reviewed these subreddits as well as the “related subreddits” listed on their community pages. We identified 47 subreddits that met the following criteria: 1) The subreddit is among the top 5\% on Reddit in terms of size, which indicated a large community; 2) The subreddit had a post or activity within the last week, which indicated an active community. After reviewing their policies, we reached out to their moderation teams to seek permission to post recruitment messages. Of these, 13 did not respond and 24 explicitly rejected our request, citing a preference to keep their community free from such activities.

Ten subreddits that permitted recruitment focused on four topical areas: six subreddits related to essential needs and financial assistance; two subreddits related to homelessness; one related to parental loss; and one related to unemployment. None of these subreddits impose minimum karma requirements, allowing new users to post. None place formal restrictions on how requests for assistance must be presented, although they advise community members to remain cautious of potential scams.

Our recruitment message included a pre-screening survey with questions about demographics, Reddit usage, and experiences seeking support. We received 293 responses to the survey. 
As we could not interview all respondents, we drew on the Social Vulnerability Index (SVI) \cite{flanagan2011social} to prioritize participants more likely to face resource constraints that could increase exposure to harm and limit their capacity to cope with adverse outcomes. We reviewed SVI-related indicators included in our survey and prioritized U.S. respondents who met at least one of the following criteria: an annual household income below \$30,000 \cite{shrider2023poverty}, unemployment, no education beyond high school, or membership in a racial or ethnic minority group.  We invited 86 respondents to schedule interviews; 25 accepted the invitation and participated in interviews. Participant demographics are provided in Table \ref{participantsTable}.
 
\begin{table}
\caption{Interview participants' demographic information and type of support sought}
\footnotesize
\begin{tabularx}{\textwidth}{p{0.3cm}p{2.3 cm}p{3cm}p{1.3 cm}p{2.3cm}p{3.5cm}}
\toprule
ID & Gender & Ethnicity & Age & Reddit usage & Support Type\\
\midrule
P1 & Female & Black or African American & 25 to 34& More than 3 years & Emotional, Informational, Tangible\\
P2 & Female & Black or African American & 25 to 34 & 1-3 years & Emotional, Informational, Tangible\\
P3 & Female & Black or African American & 18 to 24 & 1-3 years & Emotional, Informational\\
P4 & Female & Black or African American & 25 to 34 & More than 3 years & Emotional, Informational, Tangible\\
P5 & Male & Black or African American & 18 to 24 & 1-3 years & Emotional, Informational, Tangible\\
P6 & Male & Black or African American & 18 to 24 & 1-3 years & Emotional, Informational, Tangible\\
P7 & Female  & Black or African American & 18 to 24  & 1-3 years & Emotional, Informational\\
P8 & Non-binary & Black or African American & 18 to 24 & More than 3 years & Emotional, Informational, Tangible\\
P9 & Male & Black or African American & 25 to 34 & More than 3 years & Tangible\\
P10 & Male & Black or African American & 25 to 34 & More than 3 years  & Emotional, Informational, Tangible\\
P11 & Male & Black or African American & 25 to 34 & 1-3 years & Emotional, Tangible\\
P12 & Female & White & 25 to 34 & More than 3 years & Informational, Tangible\\
P13 & Female & White & 25 to 34 & More than 3 years & Emotional, Informational, Tangible\\
P14 & Male & Black or African American & 18 to 24 & 1-3 years & Emotional, Informational\\
P15 & Male & Black or African American & 18 to 24 & 1-3 years & Emotional, Informational\\
P16 & Non-binary & Black or African American &25 to 34 & 1-3 years & Emotional, Informational\\
P17 & Female & Asian & 25 to 34 & More than 3 years & Emotional, Informational, Tangible\\
P18 & Female & Asian & 18 to 24 & More than 3 years & Emotional, Informational, Tangible\\
P19 & Female & Black or African American & 25 to 34  & More than 3 years & Emotional, Informational\\
P20 & Non-binary & Black or African American & 25 to 34 & More than 3 years & Emotional, Informational\\
P21 & Female & Black or African American & 45 to 54 & More than 3 years & Emotional, Informational, Tangible\\
P22 & Non-binary & Black or African American & 25 to 34 & More than 3 years & Emotional, Informational, Tangible\\
P23 & Female & White & 18 to 24 & More than 3 years & Emotional, Informational, Tangible\\
P24 & Male & Black or African American & 18 to 24 & More than 3 years & Emotional, Informational\\
P25 & Female & Black or African American & 25 to 34 & More than 3 years & Emotional, Informational\\
\bottomrule
\end{tabularx}
\label{participantsTable}
\end{table}

\subsection{Interview Structure}
Interviews typically lasted 45-60 minutes and were conducted and recorded using Zoom teleconferencing software. Given the sensitive nature of the research topic, during the consent procedure, we assured participants that their participation was completely voluntary and that they could stop at any time. After the interview, we gave participants an opportunity to have their interview data excluded from analysis and publication; none requested exclusion. The interview began with demographic questions (e.g., gender, race, age, education, and income). We then asked about participants’ general experiences on Reddit (e.g., time spent on the platform, most frequently visited subreddits and typical interactions with other users). We invited participants to discuss specific experiences of seeking support online (e.g., what they posted, why they chose to post on Reddit). 
These could be experiences on the subreddit through which they were recruited as well as experiences on other subreddits. We then invited them to share experiences where they encountered comments that made them feel uncomfortable or unsafe and how they responded. Our complete interview protocol is provided in the Appendix. Each participant received a USD 30 Amazon gift card as a token of appreciation. The study was reviewed and approved by our university’s Institutional Review Board (IRB).

\subsection{Data Analysis}
Interview recordings were automatically transcribed into text by Zoom and reviewed manually for accuracy. We analyzed the data using a process of structural coding and subcoding \cite{saldana_coding_2013}. The first author read through each transcript and took notes to identify the ideas discussed in each interview. Informed by this familiarization process and guided by the study's research questions and interview guide, the first author developed a codebook and discussed it with the research team. The codebook included codes such as types of support, seeking support, providing support, online harm experiences, impact of harm, and platform features. The first author then conducted a round of structural coding to apply the codebook to the transcripts, followed by a round of subcoding to develop more granular codes within each structural code. For instance, the structural code on seeking support included subcodes on reasons for seeking support online, reasons for choosing Reddit, ways that people sought support, unexpected responses people received, and reactions to receiving support. The codes were inductively developed except for the types of support code, whose subcodes (i.e., emotional support, informational support, and tangible support) were directly informed by prior literature. The research team met weekly to review the coding, synthesize the main points from each code, and distill the takeaways into findings that address the research questions, resolving the minimal disagreements that arose through discussion and consensus.

\section{Findings}

To address our research questions, this section first describes how vulnerable individuals perceive harm encountered during online support seeking and how these experiences shape their ongoing engagement within online support communities. We then explore how participants interpreted, negotiated, and responded to moments of harm in relation to their self-disclosure and support-seeking practices.

\subsection{Seeking Support Under Constraints}

Participants reported seeking online support in response to a variety of significant life challenges, including marital infidelity, difficulties in interpersonal relationships, domestic violence, the process of coming out to family and friends, experiences of discrimination, job loss, serious illness requiring urgent surgery, and homelessness. Many described turning to Reddit after exhausting other avenues for support. They considered it a decision made not out of preference but out of necessity. When asked why they turned to Reddit, participants often described a lack of accessible alternatives in their offline lives. For example, P1 was a victim of domestic violence, but she said that no one in her offline social network offered her support amidst the abuse. Instead, she said they told her she should be a good wife. So, she turned to Reddit and told her story.

\begin{quotation}
``\emph{It was really a struggle... I couldn't find anyone who could understand me. I didn't know what to really do. I was blaming myself. I thought the problem was on me. I just wanted to share this with the community members.  And I don't think I can suffer.}'' -P1
\end{quotation}

In particular, most participants said they chose Reddit over other social media platforms since, among the platforms they were familiar with, Reddit’s pseudonymity provided them with a protective sense of distance. This pseudonymity shielded participants from concerns of identity exposure, allowing them to speak freely without fear of gossip, judgment, or privacy breaches. 

\begin{quotation}
``\emph{Reddit's good because the users are anonymous like myself. I'm anonymous to the users, so it is good to ask for help. They won't know. This is me... They won't judge as much.}'' -P8
\end{quotation}


Reddit occupied a simultaneously isolating and connective position, estranging participants from familiar networks while standing as the only platform widely recognized enough to feel reachable and accessible to others in similar situations. This distance from familiar networks appeared to open up, rather than foreclose, a different kind of connection: even though users interacted anonymously, participants described perceiving a form of authenticity in others' narratives that they rarely encountered on other social media platforms. For P7, this distinction arose from the pressure on other platforms to construct a desirable public image:

\begin{quotation}
``\emph{It’s more of a true reflection of what is in my real life. It is not false. People pretend what they want others to see. But behind the cutting stuff it's really different. So for Reddit, when someone posts something for emotional support, it is reality. They are open to themselves that they are going through this thing. They need help.}'' -P7
\end{quotation}

This sense of authenticity did not depend on knowing who the narrator was or verifying the factual details of their account. Instead, it emerged through subtle narrative cues, such as the use of similar words and descriptions, the articulation of familiar emotions and inner struggles, and details that resonated with participants' own experiences. As P2 explained:

\begin{quotation}
``\emph{Usually, I’m not sure who’s behind the screen, of course, but these are experiences that would usually be exclusive to someone like me. The way they would know or the experiences with such things, or such situations. And you're like, it's only people like this who experience this in this kind of way.}'' -P2
\end{quotation}

Recognizing these traces of shared feeling allowed participants to see themselves in others' stories and to experience anonymous strangers as people who genuinely understood what they were going through.

Participants described receiving various forms of support, including emotional, informational, and tangible support. Responses that express emotional support provide empathy, caring, reassurance, trust, and  opportunities for emotional expression and venting \cite{langford1997social}. For example, P1 said that people on Reddit ``\emph{told me not to really blame myself for it... I had a lot of voices, and I feel those built me,}'' reflecting that strangers’ validation helped her momentarily regain strength. Informational support typically takes the form of advice or guidance in dealing with one’s problem \cite{langford1997social}. Participants experienced informational support when responses provided relevant information intended to help participants cope with current difficulties. For example, P10 noted informational support on Reddit as a way of learning how to locate and engage with appropriate resources, even when no direct solution to their specific problem was available.

\begin{quotation}
``\emph{Maybe you have a challenge, and you don't know how to go about it. Yeah, from the group you can be able to get information that can link you to the exact platform where you can be able to go and get that support.. But from the group you got the information you'd be able to go there. And from there you're able to get whatever support you want.}'' -P10
\end{quotation}

Tangible support refers to the provision of material aid, such as financial or material support \cite{langford1997social}. Many participants sought tangible support. This  ranged from something as small as covering the cost of a subway fare to attend a job interview after unemployment or sending a pizza in winter, to something as large as paying for a surgery or providing a family with six months' worth of food. 

\begin{quotation}
``\emph{I went through a serious medical condition [and asked] for financial support from some people who were willing. In this case, I was on the receiving end for food and some basics.}'' -P6
\end{quotation}

Through these interactions, participants found the limited yet crucial help that allowed them to keep moving forward in the face of hardship.

\subsection{Perceiving Harm as Loss of Narrative}
Across interviews, participants shared a common understanding that vulnerability was both the precondition and the price of seeking support on Reddit. Disclosing one’s struggles was essential to elicit empathy, yet it simultaneously exposed them to harm. 

\begin{quotation}
``\emph{Like the really vulnerable stuff, maybe about sexuality or something I'm struggling with mentally, and now the struggles with my body and my security is out there...that makes me a target, a target for people to bully me...that I'm giving too much information about the negative part of myself... makes me feel weak or sad.}'' -P2
\end{quotation}

Participants viewed online harm not as an unforeseen accident, but as an anticipated risk embedded in the act of disclosure. Most participants repeatedly emphasized that “\emph{not everyone online is kind},” and described harm as a foreseeable threat rather than a sudden shock. Many developed defensive emotional strategies and reminded themselves that, as P3 said, “\emph{if I put [negative comments] in my heart, it will affect me for a while. I have to ignore it.}” This self-protective posture, while adaptive, did not equate to immunity; the constant anticipation of hostility carried psychological costs, gradually eroding participants’ trust in others and their sense of self-worth.

Participants emphasized that online harm extends beyond the mere use of offensive or derogatory language---the kind of harm that Reddit’s current word-filtering tools are designed to detect. Rather than focusing on explicit insults, participants described harm as emerging from the intent and tone of interaction, whether they sensed malice in others’ responses. 

\begin{quotation}
``\emph{You can clearly see I’m black and everything. Then you make it an issue... Like, you create an issue. When you read the comment, you can tell this definitely. This person has an issue with you and everything. With your color. With your skin. With everything.}'' -P14
\end{quotation}

As P14 noted, the harm they experienced was rooted in the dynamics of interaction with readers or audiences, in other words, how others in the subreddit reacted to, reframed, or undermined their disclosures. We describe this as a loss of narrative, where participants' personal stories or experiences are misunderstood, dismissed, judged, or distorted by perpetrators. Below, we describe three dimensions of narrative loss that participants experienced through online support seeking.

\subsubsection{Shifting Narrative Authority}
Participants’ experiences revealed that once their stories entered Reddit’s public sphere, interpretive authority shifted from the narrator to the audience. To gain trust within the community, many felt compelled to disclose extensive personal details. But these disclosures did not guarantee that audiences would interpret their experiences as intended. Posts intended to express distress were frequently reinterpreted or dismissed by others. For instance, P6 asked for food and basic necessities during his bankruptcy and homelessness. After several commenters publicly challenged his credibility, saying things like ``\emph{don’t believe him}'' and ``\emph{what you’re saying is not true,}'' the support in the thread ceased altogether.

In other cases, readers reduced participants’ multifaceted struggles to narrow, stigmatizing interpretations. P4, who was dealing with health concerns, financial instability, and safety risks in her living environment, turned to Reddit for informational support. Instead of engaging with what she asked for, commenters substituted their own explanation for her circumstances, centered on her weight and socioeconomic status, and responded with judgment in place of assistance.

\begin{quotation}
``\emph{I thought most of the people would see the story from my perspective. But that's the... as most of the people from the other perspective, the other negative perspective and I really got a lot of backlash from it I really got a lot of backlash from it, or whatever you put it. I mean, everyone has different opinions. Something will come to you, like you must be ugly. You must be body fat. Telling me to just to not be fat.}'' –P4
\end{quotation}

The audience displaced the account she had offered—one connecting health, financial, and environmental insecurity—and substituted a story of individual bodily and moral failure. Similar situations appeared across other participants. Women who shared experiences of domestic violence were told that their personalities had provoked the abuse; homeless individuals were accused of deserving hunger as they are poor; and LGBTQ participants facing discrimination were told they could avoid mistreatment if they concealed their identities. These examples collectively reveal how online audiences often reduce complex life circumstances and personal suffering into simplistic causal judgments, erasing the nuances and emotions that participants sought to convey.  They imposed a different narrative structure: responsibility was reassigned, structural and relational conditions were removed. The online self behind the text constructed through one's experiences, along with one's thoughts and emotions, was rewritten by third-party commenters into an entirely different story. Their narratives were redefined by audiences and stripped of their original intent. 
The harm extended beyond verbal insult. It reflected a deeper deprivation of discursive power, telling participants, “Your account of your own experience is wrong.”

\subsubsection{Fragmenting Narrative Coherence}
Another dimension of narrative loss that participants experienced was a loss of narrative coherence. What was originally shared as a coherent experience, meant to make sense of pain or seek guidance, became fragmented through selective reading, hostile responses, or dismissive interpretations. Most participants noted users would read only part of a thread and then ``\emph{jump in}'' to leave a controversial comment. 

For example, P25 described experiencing racial profiling, where others made identity judgments without evidence. Although she deliberately avoided revealing her race or nationality and considered such details irrelevant to her life challenge, readers inferred that she was Black based on selective narratives in her post and responded through biased assumptions and stereotypes. The thread quickly devolved into a racial debate, shifting attention away from her original concern and leaving her request for support unanswered.

\begin{quotation}
``\emph{Some of them even go off to pick and see things, somewhat like, I'm sure you are black. Only black people do use it. So when such certain topics come up, it cuts me off guidance.}'' –P25
\end{quotation}

P9 shared a similar experience where his story was dissected into fragments and stripped of meaning. He said he came to Reddit because “I just needed a way out, that’s why I was trying to express myself.” But instead of empathy, he encountered judgment and criticism that made him feel irredeemable.

\begin{quotation}
``\emph{I was trying to explain...very detailed about my experience. All the events that actually happens to me. I've made the mistakes already. So we don't need to be finding faults. We should be looking for solutions. But when someone seems to be focusing more on my fault. Things like these make me feel less of myself.}'' –P9
\end{quotation}

The  platform that enabled him to share his story also amplified other voices that overshadowed his own. The space that was meant to provide expression and relief ultimately undermined both.  
Previous literature has found that self-disclosure is a process that allows people to reflect on their experiences and integrate trauma \cite{luo2020self}. However, our participants' experiences suggest that disruptions of narrative can lead to self-doubt and denial in support seeking. 

\subsubsection{Shrinking Narrative Space}
We also observed participants experiencing a loss of narrative space, where they felt increasingly constrained in the places and ways in which they were allowed to tell their own stories. Some participants experienced this constraint as a form of exclusion, a sense of being pushed out of spaces where their experiences could be expressed and recognized. Only narratives that aligned with the community’s implicit “victim script” were granted sympathy or support; once a story deviated from these expectations, it was met with disbelief, hostility, or outright rejection. For instance, several LGBTQ+ participants described turning to online spaces precisely because they lacked similar connections offline, only to find those same spaces unwilling to recognize their identities. P8, who identified as non-binary, described writing diary-like posts about their personal experiences and feelings. Through these posts, they hoped to find resonance with others. Instead of engaging with the experiences P8 described, however, some users questioned whether P8 genuinely belonged to the community:

\begin{quotation}
“\emph{When I posted about my story, someone said that you guys love using non-binary people in the writing because you are not part of the community.}” -P8
\end{quotation}

P8 explained that this comment meant that other users treated their account as evidence that they were not authentically part of the community. P8 had entered the space hoping to learn whether other community members had felt or experienced something similar, but the response shifted attention away from those experiences and toward whether they were entitled to speak as a non-binary person. As P8 further reflected:

\begin{quotation}
“\emph{Yes, they are insecure. I am insecure. But we own different experiences. They are insecure about themselves, but they are projecting it on you.}” -P8
\end{quotation}

Participants perceived harm from such exclusion not only because of the denial of their identity, but also because their disclosure had to conform to a predetermined script in order to be considered real and to receive support.

Loss of narrative space also manifested in participants' internalized withdrawal. Many participants described being driven out of online support by their own fear. This can be understood as self-silencing, which refers to the deliberate restraint of self-expression as a means of emotional self-protection or avoidance of potential conflict \cite{harper2006self,10.1145/3701191}. For example, participant P11 described how discussions he felt were related to his identity often shifted from a friendly atmosphere to hostile attacks, so he decided to avoid posts containing those keywords. As he explained, suppressing his own voice came at a cost:

\begin{quotation}
“\emph{Talking less is kind of a disadvantage, because sometimes you have some specific things that you really want to share, or you want people to give insights on, or you want people to know about. But you cannot share, because you're scared.}” -P11
\end{quotation}

Participants’ capacity to speak becomes conditioned by what the community deems acceptable or worth responding to. When support requires conformity to dominant narratives of vulnerability, those whose identities or experiences diverge find their stories gradually pushed to the margins. These experiences show that overt malice and social, emotional, and structural pressures render some support seekers' voices less sustainable than others. Together, the elements of exclusion and self-silencing constructed a shrinking narrative space for participants.

\subsection{Negotiating Vulnerability and Limited Agency}
Participants described online support seeking as a continual negotiation between visibility and safety, disclosure and restraint. Their agency was characterized by carefully managed vulnerability under constraints. Experiencing harm fundamentally reshaped how participants approached online interaction, forcing them to develop defensive, strategic ways of navigating care within risk. 

Participants’ risk perceptions stemmed from their experiences of being harmed in the online environment. As noted in the previous section, online harm is often subtle and unpredictable. Its subtlety lies not merely in linguistic aggression but in how it deprives support seekers of already limited resources. The harm operates through disempowerment rather than overt hostility, stripping individuals of agency and dignity in spaces where they seek support. Victims and even those who share similar experiences are hurt by these comments. However, as this content seldom contains explicit toxic terms, it bypasses moderation systems.

\begin{quotation}
``\emph{If they feel like you're doing something that's against their rules. They'll remove your post. But, you can just do something, you're within the rules. Even if others report you, the moderators are not a hundred percent sure, they're okay, but not fully. Negative comments will be still thrown out there.}'' -P16
\end{quotation}

Unpredictability also stemmed from the fact that support seekers often cannot identify or predict who a perpetrator might be. As P17 explained, perpetrators sometimes pose as empathetic listeners to elicit personal details before turning hostile, weaponizing disclosure itself:

\begin{quotation}
``\emph{But sometimes you find that someone has commented to you before. and you say, Hi, you think like, it's a nice person. Then, later on, when they start, those bad comments. You never knew that they were going to be like that. There is scary, because of this, I never know who is going to be bad and who is going to be good.}'' -P17
\end{quotation}

These experiences reflect asymmetrical power. Platform features such as pseudonymity and asynchronism allow others to distort, mock, or dismiss a person’s story without accountability, leaving support seekers to absorb harm unilaterally. In such asymmetrical encounters, vulnerability becomes both a precondition for connection and a site of exploitation. Although support seekers occupy a disadvantaged position, they are not passive victims. Rather, they actively employ a range of defensive and reparative strategies to navigate between risk and support, striving to secure a minimal but meaningful sense of agency. For many participants, agency becomes an ongoing negotiation: deciding what to disclose, and how to remain safe without withdrawing entirely. We identified two strategies participants used in this negotiation process.

\subsubsection{Reconstructing Coherence and Authority: Self-Disclosure Management}
While self-silencing can protect a support seeker from further harm, it comes at the cost of sacrificing the core goal of seeking support. When participants still needed support, they turned to self-disclosure management as a positive form of negotiation in an attempt to repair or rebuild the coherence and authority of the narrative of suffering in their lives.

One way they did this was to separate their vulnerable self from their public narrative. For example, P19 revised or removed specific details, edited sensitive information, and sometimes reframed their experiences as “a story” or “someone else’s situation.”

\begin{quotation}
``\emph{If any current situation and passing through. I just kind of made that supposed to be anonymous. It might not be shown that is actually myself I'm talking about. you know. Let me just make it like a story.}'' -P19
\end{quotation}

By sacrificing authenticity and narrative completeness, support seekers gained emotional safety and anonymity. Such defensive reconstruction allowed them to maintain a fragile sense of control—reassuring themselves that any negative feedback was directed toward a reframed version of the self, not their “real” identity.

Using throwaway accounts supported this process. As P17 explained, harm on Reddit targets a digital identity rather than the person behind it:

\begin{quotation}
``\emph{You know, let's say in Instagram. You'll post photos. Or you're using your name. I don't feel that comfortable sharing, personally, because they're looking and judging everything I'm doing. It will come back to me personally. But on Reddit, even if they comment some negative things based on what I've posted, comments will come back to me, but not directly to me. If my name is rabbits, or that rabbit's girl, it will come to that rabbit girl, but no one no one like come to me unless you tell someone.}'' -P17
\end{quotation}

Participants reinstated a boundary between their digital and offline selves. This was also a boundary that isolated harm and maintained the integrity of participants' core identity. When authority over a primary account became untenable, participants established disposable, temporary forms of authorship that enabled their continued participation while shielding the authentic self.

\subsubsection{Reclaiming the Narrative: Collective Interventions and Rebuilding Norms}
Participants not only adapted their individual behavior to protect themselves, but also attempted to reshape community norms to prevent online harm.

Some participants said that when the majority of a community shares the same attitude toward online harm, even if a newcomer with malicious intent enters the community, their harmful behavior is often interrupted and corrected by the established community. So, when they observed harmful content in the community, they actively commented below, showing their support for the poster. 

\begin{quotation}
``\emph{I feel like, if everyone is against the whole being mean to people and stuff, a few people. It will be rare to get like people who are mean to others. because if the community is anti-bullying. they won't be paying for anyone to bully anyone.}'' -P4
\end{quotation}

A few participants from the LGBTQ+ community, for example P20, shared how they corrected others' prejudiced perceptions about LGBTQ+ groups and engaged in the process of educating the public.

\begin{quotation}
``\emph{Seeing that person apologized. It means that he's not really at fault, it is just because they don't really know things about us. That's why they choose to act and say those sorts of things [gender transition]. Now we did see this thing because it's no harm.}''  -P20
\end{quotation}

They received apologies from those individuals, which P20 saw as a beneficial approach for other groups or the majority to truly understand them. They expected that this would reduce misunderstandings and hostilities from other groups.

Others created alternative infrastructures such as external mutual aid networks, offering private emotional or material support beyond Reddit’s public visibility. P5 organized such an initiative:

\begin{quotation}
``\emph{We run the online mutual service team. We just provide help for them, so that they can also feel they are valued or cared for. And we just make them feel like just forget about what happened.}'' -P5
\end{quotation}

When the self-organized mutual service team noticed some users facing a large number of negative comments, they would privately message that user and engage in deep conversations to make them feel cared for and supported by others.

Through these small-scale interventions, participants transformed individual vulnerability into relational and collective resilience. They protected fragments of selfhood, reconstructed broken narratives, and reshaped communal norms to make space for care.

\section{Discussion}
Through interviews with vulnerable Reddit users, we found that seeking support online is a constrained and negotiated process. Participants' efforts to seek support online were shaped by scarcity and isolation, and  self-disclosure became both a prerequisite and a risk. 

\subsection{Narrative Harm}
We find that when people post in online support communities, they construct and disclose an account of who they are, what has happened to them, and what they need. Such actions involve selectively organizing experiences, emotions, and circumstances into what participants themselves described as a “story.” Although grounded in lived experience, these stories sometimes omitted, altered, or even fabricated details to protect participants’ privacy while preserving what they regarded as their experiential truth. Prior HCI research, drawing on dramaturgical theory \cite{goffman2023presentation,hogan2010presentation}, has examined online self-presentation as the strategic management of images, text, and other social cues to perform particular roles or produce desired impressions \cite{zhao2013many,pinch2024subtleties,gran2025performing}. Our findings foreground another dimension of authenticity: participants valued whether a story faithfully conveyed what their experiences meant and felt like. Recognition and resonance arose when others understood the self articulated through that story. Drawing on narrative identity theory \cite{mcadams2013narrative}, we therefore conceptualize storytelling as both communicative and constitutive. Narrative not only conveys information and facilitates support seeking; it also provides a medium through which individuals organize experience, interpret their lives, and articulate who they are \cite{ricoeur1980narrative,ricoeur1991human,mclean2015personal}.

Audiences do not merely receive a narrative. Their responses shape which interpretations are recognized and whether the story remains tellable. This leads us to ask: when and how is a narrator’s agency to shape, sustain, and continue telling their own story undermined? 

It is from this understanding of support seeking as a fundamentally narrative process that we theorize narrative harm: a form of harm that occurs when an audience's interpretation of a first-person account overtakes and displaces the narrator's own, undermining their standing to shape, sustain, and continue telling their own story. Asynchronism in online communities can exacerbate this harm. Unlike in live or face-to-face interaction, the original poster cannot immediately repair misunderstandings, redirect the conversation, or alert authorities, so audience interpretations can circulate and solidify before the narrator is able to respond.

The loss of narrative coherence, interpretive authority, and the space for voice constrain agency in story construction: a support seeker's practical capacity and willingness to formulate, sustain, revise, and continue disclosing own story in support-seeking interactions. This surfaced directly in our interviews, where participants asked researchers whether they had done something wrong or whether they should stop voicing altogether. We treat those actions as a form of loss consistent with harm as a setback to one's fundamental interests, well-being, or rights \cite{feinberg1989moral}. 

Narrative harm is not the same as negative reaction or disagreement. Audiences inevitably interpret a story from perspectives that differ from the narrator's own, and such differences do not by themselves constitute harm. Narrative harm occurs when an audience's interpretation becomes dominant enough to displace the narrator's account as the organizing frame of the interaction, when the narrator is present, has disclosed, and has been read, yet no longer retains meaningful authority over what that disclosure is taken to mean.

Narrative harm's focus on individual agency distinguishes it from several related concepts. Epistemic injustice broadly concerns wrongs to people as knowers, and testimonial injustice more specifically describes credibility deficits from identity-based prejudice \cite{fricker2017evolving}. Narrative harm may involve such judgments, but neither disbelief nor prejudice is necessary for harm to occur: an account can be believed and still have its meaning appropriated around the audience's preferred explanation. Similarly, audiences may enact representational harm \cite{suresh2021framework, shelby2023sociotechnical, shelby2023sociotechnical} by reducing a support-seeker's story to a stereotypical representation of a group. Representational harm addresses the issue of how systems or society misrepresent groups, whereas narrative harm addresses how online interactions can misrepresent or weaponize people's own stories, stripping them of the right to narrate their own story.

Stigmatization concerns the devaluation of people through socially marked categories \cite{link2001conceptualizing}. Stigmatizing interpretations can produce narrative harm, but displacement can also occur through individualized blame or moral judgment without invoking a stigmatized category at all. Invalidation similarly involves rejecting another's subjective experience, but narrative harm centers on how such responses affect the narrator's continuing authority and capacity to sustain an account. Conversational exclusion \cite{baumeister1995need,williams2007ostracism} is not required for narrative harm to arise. A support seeker may remain visibly present and continue participating while audience responses crowd out, redirect, or dominate the meaning of their disclosure.

\subsection{Pseudonymity Redistributes Interpretive Power}

Reddit’s design prioritizes openness and unrestricted expression \cite{singer2014evolution}. The harms we identify are not a platform “glitch,” but a consequence of its core logic functioning as intended. In support-seeking contexts, however, these dynamics come into tension with vulnerable users’ needs for narrative coherence and interpretive authority. Anonymity and pseudonymity are key design features of Reddit \cite{andalibi2019happens,de2014mental,andalibi2017sensitive}. Past studies highlighted the disinhibition effect of anonymity and pseudonymity \cite{lapidot2012effects,10.1145/3701210}. That is, anonymity and pseudonymity can facilitate the sharing of vulnerable experiences but can also make individuals more prone to engaging in malicious behaviors. Consistent with this, our participants were aware that seeking support in pseudonymous online environments entailed risks. 

Our findings further suggest a structural consequence that extends beyond these anticipated risks and has been underexplored. While pseudonymity protects users’ identities, it also weakens their ability to defend the authority of their own narratives. In identifiable or offline contexts, individuals can rely on social identity, relational history, and community standing to contest misinterpretations and establish credibility. Under pseudonymity, these resources are largely unavailable. As a result, other users can reinterpret, question, or dismiss disclosures without accountability or contextual knowledge, while posters have limited capacity to respond. Thus, pseudonymity simultaneously enables disclosure and undermines users’ control over their narratives. As the conditions for being heard, understood, and validated are systematically undermined, harm extends beyond negative interpersonal interactions to the erosion of users' ability to meaningfully participate in support-seeking processes. The harm stems from the double-edged pseudonymity design feature and intersects with the vulnerabilities that users disproportionately carry.  Contrary to the idea that online anonymity/pseudonymity may mitigate offline power asymmetries, our findings suggest that pseudonymity may redistribute interpretive power in ways that disadvantage those already marginalized. Platforms should consider designing interventions to help users meaningfully regain narrative control and creating governance mechanisms that  center the needs of structurally disadvantaged support seekers.

\subsection{Constrained Access and the Burden of Safety Labor}

Prior research on online support seeking has largely framed it as a voluntary behavior shaped by user preferences, platform affordances, and community characteristics \cite{skeels2010catalyzing, de2014seeking, yang2019channel,young2019girl,tomprou2019career}. Our findings complicate this account in two key ways.

First, our participant sample focuses on individuals experiencing offline resource scarcity. For these users, online support seeking is a constrained coping strategy to address unmet needs, rather than a supplementary resource. Thus, costs of harm differ fundamentally from those faced by users with viable alternatives, as these individuals often lack meaningful recourse. The structural constraints shaping these users' support seeking are often obscured if online participation is examined without attention to offline status and contextual vulnerability. 

A substantial body of HCI research has examined and designed social computing systems that mediate connections between offline support provider networks and the communities, thereby facilitating access to offline support and resources \cite{dombrowski2013takes, kuo2023understanding, sha2025meeting}. If we conceptualize access to support presented in the participants’ experience as a pathway from the user to an offline resource, such as a shelter or food bank, our paper focuses on a more specific part of that pathway: the connection from the user, through a public online space, to a network of potential support providers. Our findings show that this pathway may break down before users can reach the relevant resources. When these users are harmed within online support channels, they are not redirected to alternative resources but instead returned to the same conditions of scarcity that initially drove them online. The disruption of access to information channels thus links narrative harm to broader dynamics of information inequality. Offline marginalization both produces reliance on what can be understood as appropriated infrastructures and heightens vulnerability to their exclusionary effects. While prior work has shown how socioeconomic and structural factors shape the digital divide \cite{zhai2022disparities}, our findings further suggest that harms in online support communities are not only issues of well-being but also manifestations of information inequality, disproportionately affecting  vulnerable populations.

Second, participants’ defensive strategies, such as reframing personal experiences as hypothetical stories, frequently creating and abandoning accounts, omitting identifying details, and remaining silent on certain topics after encountering hostility, were not experienced as empowering uses of platform affordances. Instead, they constitute ongoing and often invisible forms of labor required to manage vulnerability. This perspective connects our work to past literature  which examines how digital platforms shift the burden of safety onto their users \cite{o2025online}. This raises a core normative question in platform design: the challenge lies not in how to provide vulnerable users with better tools, enabling them to engage in more rational self-disclosure or employ support-seeking strategies to manage risks on their own, but rather in how to integrate the burden of safety maintenance, currently borne by users, into the platform's own design and infrastructure.

\subsection{Design Implications}
Platform design always involves trade-offs. A public forum originally intended for open discussion might be repurposed for online support, demonstrating that the same features can yield different outcomes depending on the context of use. The question is how, and by whom, the costs of communication should be borne. Our findings suggest that some costs are rendered invisible and disproportionately absorbed by support seekers who are already in vulnerable circumstances.

Rather than treating these trade-offs as fixed consequences of platform-wide defaults, we argue that they should be made visible, explicitly communicated, and open to contextual design and configuration. In other words, users and communities should at minimum be informed of what a particular design prioritizes, what costs it may introduce, and when alternative configurations are available for different interactional contexts. This is not to suggest that design can eliminate the underlying tensions, but that it can relocate them from opaque defaults into more transparent and deliberate sites of choice. Guided by this position, we propose the following design implications for enhancing user's agency and supporting context-sensitive forms of online support seeking.

\subsubsection{Maintaining Narrative Space} 
Platform design should move beyond existing engagement-led sorting mechanisms to support narrative-centered feedback spaces in this context. Sorting based on upvotes or time prioritizes collective consensus and viral engagement. In support-seeking contexts, however, these design features often marginalize the poster’s voice by elevating controversial or reductive comments over supportive dialogue. We propose that systems should introduce narrative-preserving ranking criteria that prioritize coherence and authority. Instead of presenting all responses with equal prominence or based on raw popularity, the system could prioritize displaying the poster’s own follow-up clarifications and coherent statements at the top of the thread. For example, if online commenters dismiss a support-seeker's request as a scam, as happened to one of our participants, a narrative-preserving ranking system could elevate the support seeker’s responses confirming the veracity of the request in the reply order. This shift could reduce the visual and emotional weight of harmful or off-topic comments, transforming the platform from a fragmented public forum into a structured space for narrative restoration. Such tools could focus visibility toward support-seekers' posts, directing people's limited attention toward support requests. However, since such tools could affect legitimate explorations of disagreements, misinformation, or fraud, such tools should be governed at the community level rather than imposed across a platform. 

\subsubsection{Supporting Care-Oriented Norms Through Reflective Friction} 
Our findings show that participants often had to clarify their accounts, challenge dismissive interpretations, and attempt to restore care-oriented norms themselves in online support communities—even while seeking support. Platform design should not aim to make vulnerable users more responsible for managing their own safety. Instead, design should promote prosocial interactions and try to intervene when responses could escalate into narrative harm.  
As posts involving narrative harm can appear civil and compliant with subreddit rules, they
may be difficult to identify reliably through linguistic features alone. We therefore propose a non-punitive, context-triggered reflective prompt before a reply is submitted to a support-seeking post. This mechanism would not determine whether the poster has told the “right” story, assess the truth of their account, or prescribe a single form of support; rather, it could make care-oriented norms salient to responders and shift some responsibility for maintaining them from the poster to the community. Such "reflective friction" may help  respondents think twice before posting  harmful replies \cite{van2017thinking,caraban201923}. A comment interface could introduce reflective friction for respondents who are replying to support-seeking posts. For example, such an interface could remind respondents, “This person is seeking support. Before replying, consider whether your response respects their account of their experience.”

\subsubsection{Connecting to Offline Support} 
Alleviating users' distress requires more than agency within the public space; it requires connecting them to actual offline resources or reducing the constraints of their situation. While our first two design implications focus on the links between users and public online spaces, this raises the question of how to support users to move from public online spaces to offline supports. Our participants noted that even in active subreddits, many requests go unanswered. Interactive features, such as a bot that detects help-seeking posts and responds by redirecting users toward relevant information and resources, could help match posters with resources. For instance, if a support-seeker mentions experiencing hunger in a given location, a bot could respond with links to local food banks. We acknowledge that this approach carries its own risks: outdated information or malicious bots could distract users from legitimate paths to help, and an automated response may itself lead some users to abandon their intent to share information \cite{quinn2025heterogeneous}. But such features could mitigate experiences of non-response or harm by providing support-seekers with some sense of that their post is being acknowledged.


\section{Future Work}
Based on our findings, we offer recommendations for future work. Our study centers on participants’ experiences and sense-making of perceived harm during online support seeking. Future work can explore how other users or moderators perceive online requests for support, particularly in light of the potential for narrative harm. Future work can also analyze online conversations in situ to examine how processes like shifting narrative authority occur in practice, how harm is produced, perceived, and potentially contested across different actors, and how the visibility that opens support-seekers up to narrative harm can also lead to experiences of receiving support. 
Finally, future work can study how these dynamics unfold over extended periods of time. Longitudinal or ethnographic studies could illuminate how repeated exposure to narrative harm shapes individuals’ engagement with online support spaces and their longer-term trajectories of help-seeking.

\section{Conclusion}

Through 25 semi-structured interviews, this study expands current understandings of online support.  For vulnerable individuals, engagement with online support spaces is often shaped by constrained offline resources. Our findings point to subtle forms of harm that may evade moderation, including those that undermine narrative authority, coherence, and space.  Although participants demonstrated agency through defensive and adaptive strategies, these efforts imposed a heavy burden and shifted the responsibility for ensuring safety onto those with the fewest resources. We highlight the sociotechnical shortcomings of current platforms in supporting vulnerable help-seekers and call for mechanisms in such communities that redistribute discursive power and accountability in content moderation. 

\bibliographystyle{ACM-Reference-Format}
\bibliography{sample-base}


\begin{thebibliography}{153}


\ifx \showCODEN    \undefined \def \showCODEN     #1{\unskip}     \fi
\ifx \showISBNx    \undefined \def \showISBNx     #1{\unskip}     \fi
\ifx \showISBNxiii \undefined \def \showISBNxiii  #1{\unskip}     \fi
\ifx \showISSN     \undefined \def \showISSN      #1{\unskip}     \fi
\ifx \showLCCN     \undefined \def \showLCCN      #1{\unskip}     \fi
\ifx \shownote     \undefined \def \shownote      #1{#1}          \fi
\ifx \showarticletitle \undefined \def \showarticletitle #1{#1}   \fi
\ifx \showURL      \undefined \def \showURL       {\relax}        \fi
\providecommand\bibfield[2]{#2}
\providecommand\bibinfo[2]{#2}
\providecommand\natexlab[1]{#1}
\providecommand\showeprint[2][]{arXiv:#2}

\bibitem[Agha et~al\mbox{.}(2023)]%
        {agha2023strike}
\bibfield{author}{\bibinfo{person}{Zainab Agha}, \bibinfo{person}{Karla Badillo-Urquiola}, {and} \bibinfo{person}{Pamela~J Wisniewski}.} \bibinfo{year}{2023}\natexlab{}.
\newblock \showarticletitle{" Strike at the root": co-designing real-time social media interventions for adolescent online risk prevention}.
\newblock \bibinfo{journal}{\emph{Proceedings of the ACM on Human-Computer Interaction}} \bibinfo{volume}{7}, \bibinfo{number}{CSCW1} (\bibinfo{year}{2023}), \bibinfo{pages}{1--32}.
\newblock


\bibitem[Agha et~al\mbox{.}(2024)]%
        {agha2024tricky}
\bibfield{author}{\bibinfo{person}{Zainab Agha}, \bibinfo{person}{Jinkyung Park}, \bibinfo{person}{Ruyuan Wan}, \bibinfo{person}{Naima~Samreen Ali}, \bibinfo{person}{Yiwei Wang}, \bibinfo{person}{Dominic Difranzo}, \bibinfo{person}{Karla Badillo-Urquiola}, {and} \bibinfo{person}{Pamela~J Wisniewski}.} \bibinfo{year}{2024}\natexlab{}.
\newblock \showarticletitle{Tricky vs. transparent: Towards an ecologically valid and safe approach for evaluating online safety nudges for teens}. In \bibinfo{booktitle}{\emph{Proceedings of the 2024 CHI Conference on Human Factors in Computing Systems}}. \bibinfo{pages}{1--20}.
\newblock


\bibitem[Agrafiotis et~al\mbox{.}(2018)]%
        {agrafiotis2018taxonomy}
\bibfield{author}{\bibinfo{person}{Ioannis Agrafiotis}, \bibinfo{person}{Jason~RC Nurse}, \bibinfo{person}{Michael Goldsmith}, \bibinfo{person}{Sadie Creese}, {and} \bibinfo{person}{David Upton}.} \bibinfo{year}{2018}\natexlab{}.
\newblock \showarticletitle{A taxonomy of cyber-harms: Defining the impacts of cyber-attacks and understanding how they propagate}.
\newblock \bibinfo{journal}{\emph{Journal of Cybersecurity}} \bibinfo{volume}{4}, \bibinfo{number}{1} (\bibinfo{year}{2018}), \bibinfo{pages}{tyy006}.
\newblock


\bibitem[Ali et~al\mbox{.}(2024)]%
        {ali2024m}
\bibfield{author}{\bibinfo{person}{Naima~Samreen Ali}, \bibinfo{person}{Sarvech Qadir}, \bibinfo{person}{Ashwaq Alsoubai}, \bibinfo{person}{Munmun De~Choudhury}, \bibinfo{person}{Afsaneh Razi}, {and} \bibinfo{person}{Pamela~J Wisniewski}.} \bibinfo{year}{2024}\natexlab{}.
\newblock \showarticletitle{" I'm gonna KMS": From Imminent Risk to Youth Joking about Suicide and Self-Harm via Social Media}. In \bibinfo{booktitle}{\emph{Proceedings of the 2024 CHI Conference on Human Factors in Computing Systems}}. \bibinfo{pages}{1--18}.
\newblock


\bibitem[Ammari et~al\mbox{.}(2019)]%
        {ammari2019self}
\bibfield{author}{\bibinfo{person}{Tawfiq Ammari}, \bibinfo{person}{Sarita Schoenebeck}, {and} \bibinfo{person}{Daniel Romero}.} \bibinfo{year}{2019}\natexlab{}.
\newblock \showarticletitle{Self-declared throwaway accounts on Reddit: How platform affordances and shared norms enable parenting disclosure and support}.
\newblock \bibinfo{journal}{\emph{Proceedings of the ACM on human-computer interaction}} \bibinfo{volume}{3}, \bibinfo{number}{CSCW} (\bibinfo{year}{2019}), \bibinfo{pages}{1--30}.
\newblock


\bibitem[Andalibi(2019)]%
        {andalibi2019happens}
\bibfield{author}{\bibinfo{person}{Nazanin Andalibi}.} \bibinfo{year}{2019}\natexlab{}.
\newblock \showarticletitle{What happens after disclosing stigmatized experiences on identified social media: Individual, dyadic, and social/network outcomes}. In \bibinfo{booktitle}{\emph{Proceedings of the 2019 CHI Conference on Human Factors in Computing Systems}}. \bibinfo{pages}{1--15}.
\newblock


\bibitem[Andalibi and Garcia(2021)]%
        {andalibi2021sensemaking}
\bibfield{author}{\bibinfo{person}{Nazanin Andalibi} {and} \bibinfo{person}{Patricia Garcia}.} \bibinfo{year}{2021}\natexlab{}.
\newblock \showarticletitle{Sensemaking and coping after pregnancy loss: the seeking and disruption of emotional validation online}.
\newblock \bibinfo{journal}{\emph{Proceedings of the ACM on Human-Computer Interaction}} \bibinfo{volume}{5}, \bibinfo{number}{CSCW1} (\bibinfo{year}{2021}), \bibinfo{pages}{1--32}.
\newblock


\bibitem[Andalibi et~al\mbox{.}(2016)]%
        {andalibi2016understanding}
\bibfield{author}{\bibinfo{person}{Nazanin Andalibi}, \bibinfo{person}{Oliver~L Haimson}, \bibinfo{person}{Munmun De~Choudhury}, {and} \bibinfo{person}{Andrea Forte}.} \bibinfo{year}{2016}\natexlab{}.
\newblock \showarticletitle{Understanding social media disclosures of sexual abuse through the lenses of support seeking and anonymity}. In \bibinfo{booktitle}{\emph{Proceedings of the 2016 CHI conference on human factors in computing systems}}. \bibinfo{pages}{3906--3918}.
\newblock


\bibitem[Andalibi et~al\mbox{.}(2017)]%
        {andalibi2017sensitive}
\bibfield{author}{\bibinfo{person}{Nazanin Andalibi}, \bibinfo{person}{Pinar Ozturk}, {and} \bibinfo{person}{Andrea Forte}.} \bibinfo{year}{2017}\natexlab{}.
\newblock \showarticletitle{Sensitive self-disclosures, responses, and social support on Instagram: The case of \#depression}. In \bibinfo{booktitle}{\emph{Proceedings of the 2017 ACM conference on computer supported cooperative work and social computing}}. \bibinfo{pages}{1485--1500}.
\newblock


\bibitem[Anuyah et~al\mbox{.}(2023)]%
        {anuyah2023characterizing}
\bibfield{author}{\bibinfo{person}{Oghenemaro Anuyah}, \bibinfo{person}{Karla Badillo-Urquiola}, {and} \bibinfo{person}{Ronald Metoyer}.} \bibinfo{year}{2023}\natexlab{}.
\newblock \showarticletitle{Characterizing the technology needs of vulnerable populations for participation in research and design by adopting Maslow’s hierarchy of needs}. In \bibinfo{booktitle}{\emph{Proceedings of the 2023 CHI Conference on Human Factors in Computing Systems}}. \bibinfo{pages}{1--20}.
\newblock


\bibitem[Augustaitis et~al\mbox{.}(2021)]%
        {augustaitis2021online}
\bibfield{author}{\bibinfo{person}{Laima Augustaitis}, \bibinfo{person}{Leland~A Merrill}, \bibinfo{person}{Kristi~E Gamarel}, {and} \bibinfo{person}{Oliver~L Haimson}.} \bibinfo{year}{2021}\natexlab{}.
\newblock \showarticletitle{Online transgender health information seeking: facilitators, barriers, and future directions}. In \bibinfo{booktitle}{\emph{Proceedings of the 2021 CHI Conference on Human Factors in Computing Systems}}. \bibinfo{pages}{1--14}.
\newblock


\bibitem[Barak and Gluck-Ofri(2007)]%
        {barak2007degree}
\bibfield{author}{\bibinfo{person}{Azy Barak} {and} \bibinfo{person}{Orit Gluck-Ofri}.} \bibinfo{year}{2007}\natexlab{}.
\newblock \showarticletitle{Degree and reciprocity of self-disclosure in online forums}.
\newblock \bibinfo{journal}{\emph{CyberPsychology \& Behavior}} \bibinfo{volume}{10}, \bibinfo{number}{3} (\bibinfo{year}{2007}), \bibinfo{pages}{407--417}.
\newblock


\bibitem[Barta et~al\mbox{.}(2023)]%
        {barta2023similar}
\bibfield{author}{\bibinfo{person}{Kristen Barta}, \bibinfo{person}{Katelyn Wolberg}, {and} \bibinfo{person}{Nazanin Andalibi}.} \bibinfo{year}{2023}\natexlab{}.
\newblock \showarticletitle{Similar others, social comparison, and social support in online support groups}.
\newblock \bibinfo{journal}{\emph{Proceedings of the ACM on human-computer interaction}} \bibinfo{volume}{7}, \bibinfo{number}{CSCW2} (\bibinfo{year}{2023}), \bibinfo{pages}{1--35}.
\newblock


\bibitem[Baumeister and Leary(1995)]%
        {baumeister1995need}
\bibfield{author}{\bibinfo{person}{Roy~F Baumeister} {and} \bibinfo{person}{Mark~R Leary}.} \bibinfo{year}{1995}\natexlab{}.
\newblock \showarticletitle{The need to belong: desire for interpersonal attachments as a fundamental human motivation.}
\newblock \bibinfo{journal}{\emph{Psychological bulletin}} \bibinfo{volume}{117}, \bibinfo{number}{3} (\bibinfo{year}{1995}), \bibinfo{pages}{497}.
\newblock


\bibitem[B{\"a}umler et~al\mbox{.}(2025)]%
        {baumler2025towards}
\bibfield{author}{\bibinfo{person}{Julian B{\"a}umler}, \bibinfo{person}{Helen Bader}, \bibinfo{person}{Marc-Andr{\'e} Kaufhold}, {and} \bibinfo{person}{Christian Reuter}.} \bibinfo{year}{2025}\natexlab{}.
\newblock \showarticletitle{Towards Youth-Sensitive Hateful Content Reporting: An Inclusive Focus Group Study in Germany}. In \bibinfo{booktitle}{\emph{Proceedings of the 2025 CHI Conference on Human Factors in Computing Systems}}. \bibinfo{pages}{1--22}.
\newblock


\bibitem[Blank and Lutz(2018)]%
        {blank2018benefits}
\bibfield{author}{\bibinfo{person}{Grant Blank} {and} \bibinfo{person}{Christoph Lutz}.} \bibinfo{year}{2018}\natexlab{}.
\newblock \showarticletitle{Benefits and harms from Internet use: A differentiated analysis of Great Britain}.
\newblock \bibinfo{journal}{\emph{New media \& society}} \bibinfo{volume}{20}, \bibinfo{number}{2} (\bibinfo{year}{2018}), \bibinfo{pages}{618--640}.
\newblock


\bibitem[Burger et~al\mbox{.}(2025)]%
        {burger2025saying}
\bibfield{author}{\bibinfo{person}{Braeden Burger}, \bibinfo{person}{Devin Tebbe}, \bibinfo{person}{Emma Walquist}, \bibinfo{person}{Toby Kind}, {and} \bibinfo{person}{Douglas Zytko}.} \bibinfo{year}{2025}\natexlab{}.
\newblock \showarticletitle{Saying No to "Yes Means Yes": Limitations of Affirmative Consent for Mitigating Unwanted Behavior Online According to Women and LGBTQIA+ Stakeholders}. In \bibinfo{booktitle}{\emph{Proceedings of the 2025 CHI Conference on Human Factors in Computing Systems}}. \bibinfo{pages}{1--17}.
\newblock


\bibitem[Burke and Kraut(2013)]%
        {burke2013using}
\bibfield{author}{\bibinfo{person}{Moira Burke} {and} \bibinfo{person}{Robert Kraut}.} \bibinfo{year}{2013}\natexlab{}.
\newblock \showarticletitle{Using Facebook after losing a job: Differential benefits of strong and weak ties}. In \bibinfo{booktitle}{\emph{Proceedings of the 2013 conference on Computer supported cooperative work}}. \bibinfo{pages}{1419--1430}.
\newblock


\bibitem[Cai et~al\mbox{.}(2024)]%
        {cai2024content}
\bibfield{author}{\bibinfo{person}{Jie Cai}, \bibinfo{person}{Aashka Patel}, \bibinfo{person}{Azadeh Naderi}, {and} \bibinfo{person}{Donghee~Yvette Wohn}.} \bibinfo{year}{2024}\natexlab{}.
\newblock \showarticletitle{Content moderation justice and fairness on social media: Comparisons across different contexts and platforms}. In \bibinfo{booktitle}{\emph{Extended Abstracts of the CHI Conference on Human Factors in Computing Systems}}. \bibinfo{pages}{1--9}.
\newblock


\bibitem[Caraban et~al\mbox{.}(2019)]%
        {caraban201923}
\bibfield{author}{\bibinfo{person}{Ana Caraban}, \bibinfo{person}{Evangelos Karapanos}, \bibinfo{person}{Daniel Gon{\c{c}}alves}, {and} \bibinfo{person}{Pedro Campos}.} \bibinfo{year}{2019}\natexlab{}.
\newblock \showarticletitle{23 ways to nudge: A review of technology-mediated nudging in human-computer interaction}. In \bibinfo{booktitle}{\emph{Proceedings of the 2019 CHI conference on human factors in computing systems}}. \bibinfo{pages}{1--15}.
\newblock


\bibitem[Chang et~al\mbox{.}(2025)]%
        {chang2025opaque}
\bibfield{author}{\bibinfo{person}{Tyler Chang}, \bibinfo{person}{Joseph~J Trybala~III}, \bibinfo{person}{Sharon Bassan}, {and} \bibinfo{person}{Afsaneh Razi}.} \bibinfo{year}{2025}\natexlab{}.
\newblock \showarticletitle{Opaque Transparency: Gaps and Discrepancies in the Report of Social Media Harms}. In \bibinfo{booktitle}{\emph{Proceedings of the Extended Abstracts of the CHI Conference on Human Factors in Computing Systems}}. \bibinfo{pages}{1--12}.
\newblock


\bibitem[Cozby(1973)]%
        {cozby1973self}
\bibfield{author}{\bibinfo{person}{Paul~C Cozby}.} \bibinfo{year}{1973}\natexlab{}.
\newblock \showarticletitle{Self-disclosure: a literature review.}
\newblock \bibinfo{journal}{\emph{Psychological bulletin}} \bibinfo{volume}{79}, \bibinfo{number}{2} (\bibinfo{year}{1973}), \bibinfo{pages}{73}.
\newblock


\bibitem[Cutter(2012)]%
        {cutter2012vulnerability}
\bibfield{author}{\bibinfo{person}{Susan~L Cutter}.} \bibinfo{year}{2012}\natexlab{}.
\newblock \showarticletitle{Vulnerability to environmental hazards}.
\newblock In \bibinfo{booktitle}{\emph{Hazards vulnerability and environmental justice}}. \bibinfo{publisher}{Routledge}, \bibinfo{pages}{71--82}.
\newblock


\bibitem[Cutter et~al\mbox{.}(2012)]%
        {cutter2012social}
\bibfield{author}{\bibinfo{person}{Susan~L Cutter}, \bibinfo{person}{Bryan~J Boruff}, {and} \bibinfo{person}{W~Lynn Shirley}.} \bibinfo{year}{2012}\natexlab{}.
\newblock \showarticletitle{Social vulnerability to environmental hazards}.
\newblock In \bibinfo{booktitle}{\emph{Hazards vulnerability and environmental justice}}. \bibinfo{publisher}{Routledge}, \bibinfo{pages}{115--132}.
\newblock


\bibitem[Davani et~al\mbox{.}(2024)]%
        {davani2024disentangling}
\bibfield{author}{\bibinfo{person}{Aida Davani}, \bibinfo{person}{Mark D{\'\i}az}, \bibinfo{person}{Dylan Baker}, {and} \bibinfo{person}{Vinodkumar Prabhakaran}.} \bibinfo{year}{2024}\natexlab{}.
\newblock \showarticletitle{Disentangling perceptions of offensiveness: Cultural and moral correlates}. In \bibinfo{booktitle}{\emph{Proceedings of the 2024 ACM Conference on Fairness, Accountability, and Transparency}}. \bibinfo{pages}{2007--2021}.
\newblock


\bibitem[De~Choudhury and De(2014)]%
        {de2014mental}
\bibfield{author}{\bibinfo{person}{Munmun De~Choudhury} {and} \bibinfo{person}{Sushovan De}.} \bibinfo{year}{2014}\natexlab{}.
\newblock \showarticletitle{Mental health discourse on reddit: Self-disclosure, social support, and anonymity}. In \bibinfo{booktitle}{\emph{Proceedings of the international AAAI conference on web and social media}}, Vol.~\bibinfo{volume}{8}. \bibinfo{pages}{71--80}.
\newblock


\bibitem[De~Choudhury et~al\mbox{.}(2014)]%
        {de2014seeking}
\bibfield{author}{\bibinfo{person}{Munmun De~Choudhury}, \bibinfo{person}{Meredith~Ringel Morris}, {and} \bibinfo{person}{Ryen~W White}.} \bibinfo{year}{2014}\natexlab{}.
\newblock \showarticletitle{Seeking and sharing health information online: comparing search engines and social media}. In \bibinfo{booktitle}{\emph{Proceedings of the SIGCHI conference on human factors in computing systems}}. \bibinfo{pages}{1365--1376}.
\newblock


\bibitem[Derlega et~al\mbox{.}(1993)]%
        {derlega1993self}
\bibfield{author}{\bibinfo{person}{Valerian~J Derlega}, \bibinfo{person}{Sandra Metts}, \bibinfo{person}{Sandra Petronio}, {and} \bibinfo{person}{Stephen~T Margulis}.} \bibinfo{year}{1993}\natexlab{}.
\newblock \bibinfo{booktitle}{\emph{Self-disclosure.}}
\newblock \bibinfo{publisher}{Sage Publications, Inc}.
\newblock


\bibitem[Doan and Seering(2025)]%
        {doan2025design}
\bibfield{author}{\bibinfo{person}{Bich~Ngoc Doan} {and} \bibinfo{person}{Joseph Seering}.} \bibinfo{year}{2025}\natexlab{}.
\newblock \showarticletitle{The Design Space for Online Restorative Justice Tools: A Case Study with ApoloBot}. In \bibinfo{booktitle}{\emph{Proceedings of the 2025 CHI Conference on Human Factors in Computing Systems}}. \bibinfo{pages}{1--19}.
\newblock


\bibitem[Dombrowski et~al\mbox{.}(2013)]%
        {dombrowski2013takes}
\bibfield{author}{\bibinfo{person}{Lynn Dombrowski}, \bibinfo{person}{Jed~R Brubaker}, \bibinfo{person}{Sen~H Hirano}, \bibinfo{person}{Melissa Mazmanian}, {and} \bibinfo{person}{Gillian~R Hayes}.} \bibinfo{year}{2013}\natexlab{}.
\newblock \showarticletitle{It takes a network to get dinner: designing location-based systems to address local food needs}. In \bibinfo{booktitle}{\emph{Proceedings of the 2013 ACM international joint conference on Pervasive and ubiquitous computing}}. \bibinfo{pages}{519--528}.
\newblock


\bibitem[Doyle et~al\mbox{.}(2024)]%
        {doyle2024hate}
\bibfield{author}{\bibinfo{person}{Dylan~Thomas Doyle}, \bibinfo{person}{Charlie Blue~R Brahm}, {and} \bibinfo{person}{Jed~R Brubaker}.} \bibinfo{year}{2024}\natexlab{}.
\newblock \showarticletitle{" I hate you. I love you. I'm sorry. I miss you." Understanding Online Grief Expression Through Suicide Bereavement Letter-Writing Practices}.
\newblock \bibinfo{journal}{\emph{Proceedings of the ACM on Human-Computer Interaction}} \bibinfo{volume}{8}, \bibinfo{number}{CSCW1} (\bibinfo{year}{2024}), \bibinfo{pages}{1--27}.
\newblock


\bibitem[Feinberg(1989)]%
        {feinberg1989moral}
\bibfield{author}{\bibinfo{person}{Joel Feinberg}.} \bibinfo{year}{1989}\natexlab{}.
\newblock \bibinfo{booktitle}{\emph{The moral limits of the criminal law: volume 3: harm to self}}. Vol.~\bibinfo{volume}{3}.
\newblock \bibinfo{publisher}{Oxford University Press}.
\newblock


\bibitem[Flanagan et~al\mbox{.}(2011)]%
        {flanagan2011social}
\bibfield{author}{\bibinfo{person}{Barry~E Flanagan}, \bibinfo{person}{Edward~W Gregory}, \bibinfo{person}{Elaine~J Hallisey}, \bibinfo{person}{Janet~L Heitgerd}, {and} \bibinfo{person}{Brian Lewis}.} \bibinfo{year}{2011}\natexlab{}.
\newblock \showarticletitle{A social vulnerability index for disaster management}.
\newblock \bibinfo{journal}{\emph{Journal of homeland security and emergency management}} \bibinfo{volume}{8}, \bibinfo{number}{1} (\bibinfo{year}{2011}), \bibinfo{pages}{0000102202154773551792}.
\newblock


\bibitem[Forest and Wood(2012)]%
        {forest2012social}
\bibfield{author}{\bibinfo{person}{Amanda~L Forest} {and} \bibinfo{person}{Joanne~V Wood}.} \bibinfo{year}{2012}\natexlab{}.
\newblock \showarticletitle{When social networking is not working: Individuals with low self-esteem recognize but do not reap the benefits of self-disclosure on Facebook}.
\newblock \bibinfo{journal}{\emph{Psychological science}} \bibinfo{volume}{23}, \bibinfo{number}{3} (\bibinfo{year}{2012}), \bibinfo{pages}{295--302}.
\newblock


\bibitem[Fothergill and Peek(2004)]%
        {fothergill2004poverty}
\bibfield{author}{\bibinfo{person}{Alice Fothergill} {and} \bibinfo{person}{Lori~A Peek}.} \bibinfo{year}{2004}\natexlab{}.
\newblock \showarticletitle{Poverty and disasters in the United States: A review of recent sociological findings}.
\newblock \bibinfo{journal}{\emph{Natural hazards}} \bibinfo{volume}{32}, \bibinfo{number}{1} (\bibinfo{year}{2004}), \bibinfo{pages}{89--110}.
\newblock


\bibitem[Fricker(2017)]%
        {fricker2017evolving}
\bibfield{author}{\bibinfo{person}{Miranda Fricker}.} \bibinfo{year}{2017}\natexlab{}.
\newblock \showarticletitle{Evolving concepts of epistemic injustice}.
\newblock In \bibinfo{booktitle}{\emph{The Routledge handbook of epistemic injustice}}. \bibinfo{publisher}{Routledge}, \bibinfo{pages}{53--60}.
\newblock


\bibitem[Gen{\c{c}} and Verma(2024)]%
        {gencc2024situating}
\bibfield{author}{\bibinfo{person}{U{\u{g}}ur Gen{\c{c}}} {and} \bibinfo{person}{Himanshu Verma}.} \bibinfo{year}{2024}\natexlab{}.
\newblock \showarticletitle{Situating Empathy in HCI/CSCW: A Scoping Review}.
\newblock \bibinfo{journal}{\emph{Proceedings of the ACM on Human-Computer Interaction}} \bibinfo{volume}{8}, \bibinfo{number}{CSCW2} (\bibinfo{year}{2024}), \bibinfo{pages}{1--37}.
\newblock


\bibitem[Goffman(2023)]%
        {goffman2023presentation}
\bibfield{author}{\bibinfo{person}{Erving Goffman}.} \bibinfo{year}{2023}\natexlab{}.
\newblock \showarticletitle{The presentation of self in everyday life}.
\newblock In \bibinfo{booktitle}{\emph{Social theory re-wired}}. \bibinfo{publisher}{Routledge}, \bibinfo{pages}{450--459}.
\newblock


\bibitem[Goyal et~al\mbox{.}(2022)]%
        {goyal2022you}
\bibfield{author}{\bibinfo{person}{Nitesh Goyal}, \bibinfo{person}{Leslie Park}, {and} \bibinfo{person}{Lucy Vasserman}.} \bibinfo{year}{2022}\natexlab{}.
\newblock \showarticletitle{” You have to prove the threat is real”: Understanding the needs of Female Journalists and Activists to Document and Report Online Harassment}. In \bibinfo{booktitle}{\emph{Proceedings of the 2022 CHI conference on human factors in computing systems}}. \bibinfo{pages}{1--17}.
\newblock


\bibitem[Gran(2025)]%
        {gran2025performing}
\bibfield{author}{\bibinfo{person}{Anne-Britt Gran}.} \bibinfo{year}{2025}\natexlab{}.
\newblock \showarticletitle{Performing not-not-me in SoMe: a new theatrical typology of self-presentation online}.
\newblock \bibinfo{journal}{\emph{Social Media+ Society}} \bibinfo{volume}{11}, \bibinfo{number}{1} (\bibinfo{year}{2025}), \bibinfo{pages}{20563051251315256}.
\newblock


\bibitem[Griffiths et~al\mbox{.}(2011)]%
        {griffiths2011seeking}
\bibfield{author}{\bibinfo{person}{Kathleen~M Griffiths}, \bibinfo{person}{Dimity~A Crisp}, \bibinfo{person}{Lisa Barney}, {and} \bibinfo{person}{Russell Reid}.} \bibinfo{year}{2011}\natexlab{}.
\newblock \showarticletitle{Seeking help for depression from family and friends: a qualitative analysis of perceived advantages and disadvantages}.
\newblock \bibinfo{journal}{\emph{BMC psychiatry}} \bibinfo{volume}{11}, \bibinfo{number}{1} (\bibinfo{year}{2011}), \bibinfo{pages}{1--12}.
\newblock


\bibitem[Gunasekara et~al\mbox{.}(2025)]%
        {gunasekara2025influence}
\bibfield{author}{\bibinfo{person}{Suwani Gunasekara}, \bibinfo{person}{Saumya Pareek}, \bibinfo{person}{Ryan~M Kelly}, {and} \bibinfo{person}{Jorge Goncalves}.} \bibinfo{year}{2025}\natexlab{}.
\newblock \showarticletitle{The Influence of Content Modality on Perceptions of Online Misinformation}. In \bibinfo{booktitle}{\emph{Proceedings of the 2025 CHI Conference on Human Factors in Computing Systems}}. \bibinfo{pages}{1--10}.
\newblock


\bibitem[Haimson et~al\mbox{.}(2021)]%
        {haimson2021disproportionate}
\bibfield{author}{\bibinfo{person}{Oliver~L Haimson}, \bibinfo{person}{Daniel Delmonaco}, \bibinfo{person}{Peipei Nie}, {and} \bibinfo{person}{Andrea Wegner}.} \bibinfo{year}{2021}\natexlab{}.
\newblock \showarticletitle{Disproportionate removals and differing content moderation experiences for conservative, transgender, and black social media users: Marginalization and moderation gray areas}.
\newblock \bibinfo{journal}{\emph{Proceedings of the ACM on Human-Computer Interaction}} \bibinfo{volume}{5}, \bibinfo{number}{CSCW2} (\bibinfo{year}{2021}), \bibinfo{pages}{1--35}.
\newblock


\bibitem[Hardwick et~al\mbox{.}(2025)]%
        {hardwick2025they}
\bibfield{author}{\bibinfo{person}{Taylor Hardwick}, \bibinfo{person}{Marcus Carter}, \bibinfo{person}{Stephanie Harkin}, \bibinfo{person}{Tianyi Zhangshao}, {and} \bibinfo{person}{Ben Egliston}.} \bibinfo{year}{2025}\natexlab{}.
\newblock \showarticletitle{" They're Scamming Me": How Children Experience and Conceptualize Harm in Game Monetization}. In \bibinfo{booktitle}{\emph{Proceedings of the 2025 CHI Conference on Human Factors in Computing Systems}}. \bibinfo{pages}{1--14}.
\newblock


\bibitem[Harper et~al\mbox{.}(2006)]%
        {harper2006self}
\bibfield{author}{\bibinfo{person}{Melinda~S Harper}, \bibinfo{person}{Joseph~W Dickson}, {and} \bibinfo{person}{Deborah~P Welsh}.} \bibinfo{year}{2006}\natexlab{}.
\newblock \showarticletitle{Self-silencing and rejection sensitivity in adolescent romantic relationships}.
\newblock \bibinfo{journal}{\emph{Journal of Youth and Adolescence}}  \bibinfo{volume}{35} (\bibinfo{year}{2006}), \bibinfo{pages}{435--443}.
\newblock


\bibitem[Hartmann et~al\mbox{.}(2025)]%
        {hartmann2025lost}
\bibfield{author}{\bibinfo{person}{David Hartmann}, \bibinfo{person}{Amin Oueslati}, \bibinfo{person}{Dimitri Staufer}, \bibinfo{person}{Lena Pohlmann}, \bibinfo{person}{Simon Munzert}, {and} \bibinfo{person}{Hendrik Heuer}.} \bibinfo{year}{2025}\natexlab{}.
\newblock \showarticletitle{Lost in moderation: How commercial content moderation apis over-and under-moderate group-targeted hate speech and linguistic variations}. In \bibinfo{booktitle}{\emph{Proceedings of the 2025 CHI Conference on Human Factors in Computing Systems}}. \bibinfo{pages}{1--26}.
\newblock


\bibitem[Heung et~al\mbox{.}(2024)]%
        {heung2024vulnerable}
\bibfield{author}{\bibinfo{person}{Sharon Heung}, \bibinfo{person}{Lucy Jiang}, \bibinfo{person}{Shiri Azenkot}, {and} \bibinfo{person}{Aditya Vashistha}.} \bibinfo{year}{2024}\natexlab{}.
\newblock \showarticletitle{“Vulnerable, Victimized, and Objectified”: Understanding Ableist Hate and Harassment Experienced by Disabled Content Creators on Social Media}. In \bibinfo{booktitle}{\emph{Proceedings of the 2024 CHI Conference on Human Factors in Computing Systems}}. \bibinfo{pages}{1--19}.
\newblock


\bibitem[Heung et~al\mbox{.}(2025)]%
        {heung2025ignorance}
\bibfield{author}{\bibinfo{person}{Sharon Heung}, \bibinfo{person}{Lucy Jiang}, \bibinfo{person}{Shiri Azenkot}, {and} \bibinfo{person}{Aditya Vashistha}.} \bibinfo{year}{2025}\natexlab{}.
\newblock \showarticletitle{" Ignorance is not Bliss": Designing Personalized Moderation to Address Ableist Hate on Social Media}. In \bibinfo{booktitle}{\emph{Proceedings of the 2025 CHI Conference on Human Factors in Computing Systems}}. \bibinfo{pages}{1--18}.
\newblock


\bibitem[Hinchman and Hinchman(1997)]%
        {hinchman1997memory}
\bibfield{editor}{\bibinfo{person}{Lewis~P. Hinchman} {and} \bibinfo{person}{Sandra~K. Hinchman}} (Eds.). \bibinfo{year}{1997}\natexlab{}.
\newblock \bibinfo{booktitle}{\emph{Memory, Identity, Community: The Idea of Narrative in the Human Sciences}}.
\newblock \bibinfo{publisher}{State University of New York Press}, \bibinfo{address}{Albany, NY}.
\newblock


\bibitem[Hogan(2010)]%
        {hogan2010presentation}
\bibfield{author}{\bibinfo{person}{Bernie Hogan}.} \bibinfo{year}{2010}\natexlab{}.
\newblock \showarticletitle{The presentation of self in the age of social media: Distinguishing performances and exhibitions online}.
\newblock \bibinfo{journal}{\emph{Bulletin of Science, Technology \& Society}} \bibinfo{volume}{30}, \bibinfo{number}{6} (\bibinfo{year}{2010}), \bibinfo{pages}{377--386}.
\newblock


\bibitem[House et~al\mbox{.}(1988)]%
        {house1988structures}
\bibfield{author}{\bibinfo{person}{James~S House}, \bibinfo{person}{Debra Umberson}, {and} \bibinfo{person}{Karl~R Landis}.} \bibinfo{year}{1988}\natexlab{}.
\newblock \showarticletitle{Structures and processes of social support}.
\newblock \bibinfo{journal}{\emph{Annual review of sociology}} \bibinfo{volume}{14}, \bibinfo{number}{1} (\bibinfo{year}{1988}), \bibinfo{pages}{293--318}.
\newblock


\bibitem[Iivari et~al\mbox{.}(2021)]%
        {iivari2021chi}
\bibfield{author}{\bibinfo{person}{Netta Iivari}, \bibinfo{person}{Leena Vent{\"a}-Olkkonen}, \bibinfo{person}{Sumita Sharma}, \bibinfo{person}{Tonja Molin-Juustila}, {and} \bibinfo{person}{Essi Kinnunen}.} \bibinfo{year}{2021}\natexlab{}.
\newblock \showarticletitle{Chi against bullying: Taking stock of the past and envisioning the future}. In \bibinfo{booktitle}{\emph{Proceedings of the 2021 CHI Conference on Human Factors in Computing Systems}}. \bibinfo{pages}{1--17}.
\newblock


\bibitem[Jane(2015)]%
        {jane2015flaming}
\bibfield{author}{\bibinfo{person}{Emma~A Jane}.} \bibinfo{year}{2015}\natexlab{}.
\newblock \showarticletitle{Flaming? What flaming? The pitfalls and potentials of researching online hostility}.
\newblock \bibinfo{journal}{\emph{Ethics and information technology}} \bibinfo{volume}{17}, \bibinfo{number}{1} (\bibinfo{year}{2015}), \bibinfo{pages}{65--87}.
\newblock


\bibitem[Jennings et~al\mbox{.}(2014)]%
        {jennings2014intricate}
\bibfield{author}{\bibinfo{person}{Lauren~E. Jennings}, \bibinfo{person}{Monisha Pasupathi}, {and} \bibinfo{person}{Kate~C. McLean}.} \bibinfo{year}{2014}\natexlab{}.
\newblock \showarticletitle{{``Intricate Lettings Out and Lettings In''}: Listener Scaffolding of Narrative Identity in Newly Dating Romantic Partners}.
\newblock \bibinfo{journal}{\emph{Self and Identity}} \bibinfo{volume}{13}, \bibinfo{number}{2} (\bibinfo{year}{2014}), \bibinfo{pages}{214--230}.
\newblock
\href{https://doi.org/10.1080/15298868.2013.786203}{doi:\nolinkurl{10.1080/15298868.2013.786203}}


\bibitem[Jo et~al\mbox{.}(2024)]%
        {jo2024understanding}
\bibfield{author}{\bibinfo{person}{Jeongwon Jo}, \bibinfo{person}{Jingyi Xie}, {and} \bibinfo{person}{John~M Carroll}.} \bibinfo{year}{2024}\natexlab{}.
\newblock \showarticletitle{Understanding and Balancing Trade-offs of Visibility in Support Requests}. In \bibinfo{booktitle}{\emph{Extended Abstracts of the CHI Conference on Human Factors in Computing Systems}}. \bibinfo{pages}{1--8}.
\newblock


\bibitem[Jourard(1971)]%
        {jourard1971self}
\bibfield{author}{\bibinfo{person}{Sidney~M Jourard}.} \bibinfo{year}{1971}\natexlab{}.
\newblock \showarticletitle{Self-disclosure: An experimental analysis of the transparent self.}
\newblock  (\bibinfo{year}{1971}).
\newblock


\bibitem[Jourard and Lasakow(1958)]%
        {jourard1958some}
\bibfield{author}{\bibinfo{person}{Sidney~M Jourard} {and} \bibinfo{person}{Paul Lasakow}.} \bibinfo{year}{1958}\natexlab{}.
\newblock \showarticletitle{Some factors in self-disclosure.}
\newblock \bibinfo{journal}{\emph{The Journal of Abnormal and Social Psychology}} \bibinfo{volume}{56}, \bibinfo{number}{1} (\bibinfo{year}{1958}), \bibinfo{pages}{91}.
\newblock


\bibitem[Karizat et~al\mbox{.}(2025)]%
        {karizat2025laboring}
\bibfield{author}{\bibinfo{person}{Nadia Karizat}, \bibinfo{person}{Nora McDonald}, {and} \bibinfo{person}{Nazanin Andalibi}.} \bibinfo{year}{2025}\natexlab{}.
\newblock \showarticletitle{Laboring Towards Sociotechnical Reproductive Privacy in a Post-Roe United States: Identities, Technologies, and Actors Implicated in Reproductive Privacy}.
\newblock \bibinfo{journal}{\emph{Proceedings of the ACM on Human-Computer Interaction}} \bibinfo{volume}{9}, \bibinfo{number}{7} (\bibinfo{year}{2025}), \bibinfo{pages}{1--34}.
\newblock


\bibitem[Karusala and Kumar(2017)]%
        {karusala2017women}
\bibfield{author}{\bibinfo{person}{Naveena Karusala} {and} \bibinfo{person}{Neha Kumar}.} \bibinfo{year}{2017}\natexlab{}.
\newblock \showarticletitle{Women's safety in public spaces: Examining the efficacy of panic buttons in New Delhi}. In \bibinfo{booktitle}{\emph{Proceedings of the 2017 CHI conference on human factors in computing systems}}. \bibinfo{pages}{3340--3351}.
\newblock


\bibitem[Karusala et~al\mbox{.}(2021)]%
        {karusala2021courage}
\bibfield{author}{\bibinfo{person}{Naveena Karusala}, \bibinfo{person}{David~Odhiambo Seeh}, \bibinfo{person}{Cyrus Mugo}, \bibinfo{person}{Brandon Guthrie}, \bibinfo{person}{Megan~A Moreno}, \bibinfo{person}{Grace John-Stewart}, \bibinfo{person}{Irene Inwani}, \bibinfo{person}{Richard Anderson}, {and} \bibinfo{person}{Keshet Ronen}.} \bibinfo{year}{2021}\natexlab{}.
\newblock \showarticletitle{“That courage to encourage”: Participation and Aspirations in Chat-based Peer Support for Youth Living with HIV}. In \bibinfo{booktitle}{\emph{Proceedings of the 2021 CHI Conference on Human Factors in Computing Systems}}. \bibinfo{pages}{1--17}.
\newblock


\bibitem[Kim et~al\mbox{.}(2024)]%
        {kim2024respect}
\bibfield{author}{\bibinfo{person}{Haesoo Kim}, \bibinfo{person}{Juhoon Lee}, \bibinfo{person}{Jeong-Woo Jang}, {and} \bibinfo{person}{Juho Kim}.} \bibinfo{year}{2024}\natexlab{}.
\newblock \showarticletitle{ReSPect: Enabling Active and Scalable Responses to Networked Online Harassment}.
\newblock \bibinfo{journal}{\emph{Proceedings of the ACM on Human-Computer Interaction}} \bibinfo{volume}{8}, \bibinfo{number}{CSCW1} (\bibinfo{year}{2024}), \bibinfo{pages}{1--30}.
\newblock


\bibitem[Kim et~al\mbox{.}(2021)]%
        {kim2021you}
\bibfield{author}{\bibinfo{person}{Seunghyun Kim}, \bibinfo{person}{Afsaneh Razi}, \bibinfo{person}{Gianluca Stringhini}, \bibinfo{person}{Pamela~J Wisniewski}, {and} \bibinfo{person}{Munmun De~Choudhury}.} \bibinfo{year}{2021}\natexlab{}.
\newblock \showarticletitle{You Don't Know How I Feel: Insider-Outsider Perspective Gaps in Cyberbullying Risk Detection}. In \bibinfo{booktitle}{\emph{Proceedings of the International AAAI Conference on Web and Social Media}}, Vol.~\bibinfo{volume}{15}. \bibinfo{pages}{290--302}.
\newblock


\bibitem[Kojah et~al\mbox{.}(2025)]%
        {10.1145/3701191}
\bibfield{author}{\bibinfo{person}{Sena~A. Kojah}, \bibinfo{person}{Ben~Zefeng Zhang}, \bibinfo{person}{Carolina Are}, \bibinfo{person}{Daniel Delmonaco}, {and} \bibinfo{person}{Oliver~L. Haimson}.} \bibinfo{year}{2025}\natexlab{}.
\newblock \showarticletitle{"Dialing it Back:" Shadowbanning, Invisible Digital Labor, and how Marginalized Content Creators Attempt to Mitigate the Impacts of Opaque Platform Governance}.
\newblock \bibinfo{journal}{\emph{Proc. ACM Hum.-Comput. Interact.}} \bibinfo{volume}{9}, \bibinfo{number}{1}, Article \bibinfo{articleno}{GROUP12} (\bibinfo{date}{Jan.} \bibinfo{year}{2025}), \bibinfo{numpages}{22}~pages.
\newblock
\href{https://doi.org/10.1145/3701191}{doi:\nolinkurl{10.1145/3701191}}


\bibitem[Konstantinou and Karapanos(2025)]%
        {konstantinou2025behavior}
\bibfield{author}{\bibinfo{person}{Loukas Konstantinou} {and} \bibinfo{person}{Evangelos Karapanos}.} \bibinfo{year}{2025}\natexlab{}.
\newblock \showarticletitle{Behavior Change Interventions Combating Online Misinformation: A Scoping Review}. In \bibinfo{booktitle}{\emph{Proceedings of the 2025 CHI Conference on Human Factors in Computing Systems}}. \bibinfo{pages}{1--19}.
\newblock


\bibitem[Kr{\"a}mer et~al\mbox{.}(2021)]%
        {kramer2021strength}
\bibfield{author}{\bibinfo{person}{Nicole~C Kr{\"a}mer}, \bibinfo{person}{Vera Sauer}, {and} \bibinfo{person}{Nicole Ellison}.} \bibinfo{year}{2021}\natexlab{}.
\newblock \showarticletitle{The strength of weak ties revisited: Further evidence of the role of strong ties in the provision of online social support}.
\newblock \bibinfo{journal}{\emph{Social Media+ Society}} \bibinfo{volume}{7}, \bibinfo{number}{2} (\bibinfo{year}{2021}), \bibinfo{pages}{20563051211024958}.
\newblock


\bibitem[Kuo et~al\mbox{.}(2023)]%
        {kuo2023understanding}
\bibfield{author}{\bibinfo{person}{Tzu-Sheng Kuo}, \bibinfo{person}{Hong Shen}, \bibinfo{person}{Jisoo Geum}, \bibinfo{person}{Nev Jones}, \bibinfo{person}{Jason~I Hong}, \bibinfo{person}{Haiyi Zhu}, {and} \bibinfo{person}{Kenneth Holstein}.} \bibinfo{year}{2023}\natexlab{}.
\newblock \showarticletitle{Understanding frontline workers’ and unhoused individuals’ perspectives on ai used in homeless services}. In \bibinfo{booktitle}{\emph{Proceedings of the 2023 CHI Conference on Human Factors in Computing Systems}}. \bibinfo{pages}{1--17}.
\newblock


\bibitem[Lai et~al\mbox{.}(2024)]%
        {10.1145/3633068}
\bibfield{author}{\bibinfo{person}{Yu-Ju Lai}, \bibinfo{person}{Yi-Chieh Lee}, \bibinfo{person}{Chia-Chi Chang}, \bibinfo{person}{Wan-Ting Dai}, {and} \bibinfo{person}{Ying-Yu Chen}.} \bibinfo{year}{2024}\natexlab{}.
\newblock \showarticletitle{Exploring the Role of Mom's Chat Groups in the Messaging App: Enhancing Support and Empowerment for Stay-At-Home Mothers}.
\newblock \bibinfo{journal}{\emph{Proc. ACM Hum.-Comput. Interact.}} \bibinfo{volume}{8}, \bibinfo{number}{GROUP}, Article \bibinfo{articleno}{3} (\bibinfo{date}{Feb.} \bibinfo{year}{2024}), \bibinfo{numpages}{20}~pages.
\newblock
\href{https://doi.org/10.1145/3633068}{doi:\nolinkurl{10.1145/3633068}}


\bibitem[Langford et~al\mbox{.}(1997)]%
        {langford1997social}
\bibfield{author}{\bibinfo{person}{Catherine Penny~Hinson Langford}, \bibinfo{person}{Juanita Bowsher}, \bibinfo{person}{Joseph~P Maloney}, {and} \bibinfo{person}{Patricia~P Lillis}.} \bibinfo{year}{1997}\natexlab{}.
\newblock \showarticletitle{Social support: a conceptual analysis}.
\newblock \bibinfo{journal}{\emph{Journal of advanced nursing}} \bibinfo{volume}{25}, \bibinfo{number}{1} (\bibinfo{year}{1997}), \bibinfo{pages}{95--100}.
\newblock


\bibitem[Lapidot-Lefler and Barak(2012)]%
        {lapidot2012effects}
\bibfield{author}{\bibinfo{person}{Noam Lapidot-Lefler} {and} \bibinfo{person}{Azy Barak}.} \bibinfo{year}{2012}\natexlab{}.
\newblock \showarticletitle{Effects of anonymity, invisibility, and lack of eye-contact on toxic online disinhibition}.
\newblock \bibinfo{journal}{\emph{Computers in human behavior}} \bibinfo{volume}{28}, \bibinfo{number}{2} (\bibinfo{year}{2012}), \bibinfo{pages}{434--443}.
\newblock


\bibitem[Lazer et~al\mbox{.}(2018)]%
        {lazer2018science}
\bibfield{author}{\bibinfo{person}{David~MJ Lazer}, \bibinfo{person}{Matthew~A Baum}, \bibinfo{person}{Yochai Benkler}, \bibinfo{person}{Adam~J Berinsky}, \bibinfo{person}{Kelly~M Greenhill}, \bibinfo{person}{Filippo Menczer}, \bibinfo{person}{Miriam~J Metzger}, \bibinfo{person}{Brendan Nyhan}, \bibinfo{person}{Gordon Pennycook}, \bibinfo{person}{David Rothschild}, {et~al\mbox{.}}} \bibinfo{year}{2018}\natexlab{}.
\newblock \showarticletitle{The science of fake news}.
\newblock \bibinfo{journal}{\emph{Science}} \bibinfo{volume}{359}, \bibinfo{number}{6380} (\bibinfo{year}{2018}), \bibinfo{pages}{1094--1096}.
\newblock


\bibitem[Leavitt(2015)]%
        {leavitt_this_2015}
\bibfield{author}{\bibinfo{person}{Alex Leavitt}.} \bibinfo{year}{2015}\natexlab{}.
\newblock \showarticletitle{"This is a Throwaway Account": Temporary Technical Identities and Perceptions of Anonymity in a Massive Online Community}. In \bibinfo{booktitle}{\emph{Proceedings of the 18th {ACM} Conference on Computer Supported Cooperative Work \& Social Computing}} (New York, {NY}, {USA}, 2015-02-28) \emph{(\bibinfo{series}{{CSCW} '15})}. \bibinfo{publisher}{Association for Computing Machinery}, \bibinfo{pages}{317--327}.
\newblock
\showISBNx{978-1-4503-2922-4}
\href{https://doi.org/10.1145/2675133.2675175}{doi:\nolinkurl{10.1145/2675133.2675175}}


\bibitem[Lee et~al\mbox{.}(2023)]%
        {lee2023online}
\bibfield{author}{\bibinfo{person}{Jooyoung Lee}, \bibinfo{person}{Sarah Rajtmajer}, \bibinfo{person}{Eesha Srivatsavaya}, {and} \bibinfo{person}{Shomir Wilson}.} \bibinfo{year}{2023}\natexlab{}.
\newblock \showarticletitle{Online Self-Disclosure, Social Support, and User Engagement During the COVID-19 Pandemic}.
\newblock \bibinfo{journal}{\emph{ACM Transactions on Social Computing}} \bibinfo{volume}{6}, \bibinfo{number}{3-4} (\bibinfo{year}{2023}), \bibinfo{pages}{1--31}.
\newblock


\bibitem[Lee et~al\mbox{.}(2013)]%
        {lee2013lonely}
\bibfield{author}{\bibinfo{person}{Kyung-Tag Lee}, \bibinfo{person}{Mi-Jin Noh}, {and} \bibinfo{person}{Dong-Mo Koo}.} \bibinfo{year}{2013}\natexlab{}.
\newblock \showarticletitle{Lonely people are no longer lonely on social networking sites: The mediating role of self-disclosure and social support}.
\newblock \bibinfo{journal}{\emph{Cyberpsychology, Behavior, and Social Networking}} \bibinfo{volume}{16}, \bibinfo{number}{6} (\bibinfo{year}{2013}), \bibinfo{pages}{413--418}.
\newblock


\bibitem[Link and Phelan(2001)]%
        {link2001conceptualizing}
\bibfield{author}{\bibinfo{person}{Bruce~G Link} {and} \bibinfo{person}{Jo~C Phelan}.} \bibinfo{year}{2001}\natexlab{}.
\newblock \showarticletitle{Conceptualizing stigma}.
\newblock \bibinfo{journal}{\emph{Annual review of Sociology}} \bibinfo{volume}{27}, \bibinfo{number}{1} (\bibinfo{year}{2001}), \bibinfo{pages}{363--385}.
\newblock


\bibitem[Livingstone(2013)]%
        {livingstone2013online}
\bibfield{author}{\bibinfo{person}{Sonia Livingstone}.} \bibinfo{year}{2013}\natexlab{}.
\newblock \showarticletitle{Online risk, harm and vulnerability: Reflections on the evidence base for child Internet safety policy}.
\newblock \bibinfo{journal}{\emph{ZER: Journal of Communication Studies}} \bibinfo{volume}{18}, \bibinfo{number}{35} (\bibinfo{year}{2013}), \bibinfo{pages}{13--28}.
\newblock


\bibitem[Luo and Hancock(2020)]%
        {luo2020self}
\bibfield{author}{\bibinfo{person}{Mufan Luo} {and} \bibinfo{person}{Jeffrey~T Hancock}.} \bibinfo{year}{2020}\natexlab{}.
\newblock \showarticletitle{Self-disclosure and social media: motivations, mechanisms and psychological well-being}.
\newblock \bibinfo{journal}{\emph{Current opinion in psychology}}  \bibinfo{volume}{31} (\bibinfo{year}{2020}), \bibinfo{pages}{110--115}.
\newblock


\bibitem[Lyu et~al\mbox{.}(2024)]%
        {lyu2024got}
\bibfield{author}{\bibinfo{person}{Yao Lyu}, \bibinfo{person}{Jie Cai}, \bibinfo{person}{Anisa Callis}, \bibinfo{person}{Kelley Cotter}, {and} \bibinfo{person}{John~M Carroll}.} \bibinfo{year}{2024}\natexlab{}.
\newblock \showarticletitle{" I Got Flagged for Supposed Bullying, Even Though It Was in Response to Someone Harassing Me About My Disability.": A Study of Blind TikTokers’ Content Moderation Experiences}. In \bibinfo{booktitle}{\emph{Proceedings of the 2024 CHI Conference on Human Factors in Computing Systems}}. \bibinfo{pages}{1--15}.
\newblock


\bibitem[Malik and Coulson(2010)]%
        {malik2010they}
\bibfield{author}{\bibinfo{person}{Sumaira Malik} {and} \bibinfo{person}{Neil~S Coulson}.} \bibinfo{year}{2010}\natexlab{}.
\newblock \showarticletitle{‘They all supported me but I felt like I suddenly didn't belong anymore’: an exploration of perceived disadvantages to online support seeking}.
\newblock \bibinfo{journal}{\emph{Journal of Psychosomatic Obstetrics \& Gynecology}} \bibinfo{volume}{31}, \bibinfo{number}{3} (\bibinfo{year}{2010}), \bibinfo{pages}{140--149}.
\newblock


\bibitem[Maqsood and Chiasson(2021)]%
        {maqsood2021they}
\bibfield{author}{\bibinfo{person}{Sana Maqsood} {and} \bibinfo{person}{Sonia Chiasson}.} \bibinfo{year}{2021}\natexlab{}.
\newblock \showarticletitle{“They think it’s totally fine to talk to somebody on the internet they don’t know”: Teachers’ perceptions and mitigation strategies of tweens’ online risks}. In \bibinfo{booktitle}{\emph{Proceedings of the 2021 CHI Conference on Human Factors in Computing Systems}}. \bibinfo{pages}{1--17}.
\newblock


\bibitem[Masrani et~al\mbox{.}(2023)]%
        {masrani2023slowing}
\bibfield{author}{\bibinfo{person}{Teale~W Masrani}, \bibinfo{person}{Jack Jamieson}, \bibinfo{person}{Naomi Yamashita}, {and} \bibinfo{person}{Helen~Ai He}.} \bibinfo{year}{2023}\natexlab{}.
\newblock \showarticletitle{Slowing it down: Towards facilitating interpersonal mindfulness in online polarizing conversations over social media}.
\newblock \bibinfo{journal}{\emph{Proceedings of the ACM on Human-Computer Interaction}} \bibinfo{volume}{7}, \bibinfo{number}{CSCW1} (\bibinfo{year}{2023}), \bibinfo{pages}{1--27}.
\newblock


\bibitem[Mayworm et~al\mbox{.}(2024)]%
        {mayworm2024online}
\bibfield{author}{\bibinfo{person}{Samuel Mayworm}, \bibinfo{person}{Shannon Li}, \bibinfo{person}{Hibby Thach}, \bibinfo{person}{Daniel Delmonaco}, \bibinfo{person}{Christian Paneda}, \bibinfo{person}{Andrea Wegner}, {and} \bibinfo{person}{Oliver~L Haimson}.} \bibinfo{year}{2024}\natexlab{}.
\newblock \showarticletitle{The Online Identity Help Center: Designing and developing a content moderation policy resource for marginalized social media users}.
\newblock \bibinfo{journal}{\emph{Proceedings of the ACM on Human-Computer Interaction}} \bibinfo{volume}{8}, \bibinfo{number}{CSCW1} (\bibinfo{year}{2024}), \bibinfo{pages}{1--30}.
\newblock


\bibitem[McAdams and McLean(2013)]%
        {mcadams2013narrative}
\bibfield{author}{\bibinfo{person}{Dan~P. McAdams} {and} \bibinfo{person}{Kate~C. McLean}.} \bibinfo{year}{2013}\natexlab{}.
\newblock \showarticletitle{Narrative Identity}.
\newblock \bibinfo{journal}{\emph{Current Directions in Psychological Science}} \bibinfo{volume}{22}, \bibinfo{number}{3} (\bibinfo{year}{2013}), \bibinfo{pages}{233--238}.
\newblock
\href{https://doi.org/10.1177/0963721413475622}{doi:\nolinkurl{10.1177/0963721413475622}}


\bibitem[McLean and Syed(2015)]%
        {mclean2015personal}
\bibfield{author}{\bibinfo{person}{Kate~C. McLean} {and} \bibinfo{person}{Moin Syed}.} \bibinfo{year}{2015}\natexlab{}.
\newblock \showarticletitle{Personal, Master, and Alternative Narratives: An Integrative Framework for Understanding Identity Development in Context}.
\newblock \bibinfo{journal}{\emph{Human Development}} \bibinfo{volume}{58}, \bibinfo{number}{6} (\bibinfo{year}{2015}), \bibinfo{pages}{318--349}.
\newblock
\href{https://doi.org/10.1159/000445817}{doi:\nolinkurl{10.1159/000445817}}


\bibitem[Moitra et~al\mbox{.}(2020)]%
        {10.1145/3375195}
\bibfield{author}{\bibinfo{person}{Aparna Moitra}, \bibinfo{person}{Naeemul Hassan}, \bibinfo{person}{Manash~Kumar Mandal}, \bibinfo{person}{Mansurul Bhuiyan}, {and} \bibinfo{person}{Syed~Ishtiaque Ahmed}.} \bibinfo{year}{2020}\natexlab{}.
\newblock \showarticletitle{Understanding the Challenges for Bangladeshi Women to Participate in \#MeToo Movement}.
\newblock \bibinfo{journal}{\emph{Proc. ACM Hum.-Comput. Interact.}} \bibinfo{volume}{4}, \bibinfo{number}{GROUP}, Article \bibinfo{articleno}{15} (\bibinfo{date}{Jan.} \bibinfo{year}{2020}), \bibinfo{numpages}{25}~pages.
\newblock
\href{https://doi.org/10.1145/3375195}{doi:\nolinkurl{10.1145/3375195}}


\bibitem[Morrow(1999)]%
        {morrow1999identifying}
\bibfield{author}{\bibinfo{person}{Betty~Hearn Morrow}.} \bibinfo{year}{1999}\natexlab{}.
\newblock \showarticletitle{Identifying and mapping community vulnerability}.
\newblock \bibinfo{journal}{\emph{Disasters}} \bibinfo{volume}{23}, \bibinfo{number}{1} (\bibinfo{year}{1999}), \bibinfo{pages}{1--18}.
\newblock


\bibitem[Mouzannar et~al\mbox{.}(2019)]%
        {mouzannar2019fair}
\bibfield{author}{\bibinfo{person}{Hussein Mouzannar}, \bibinfo{person}{Mesrob~I Ohannessian}, {and} \bibinfo{person}{Nathan Srebro}.} \bibinfo{year}{2019}\natexlab{}.
\newblock \showarticletitle{From fair decision making to social equality}. In \bibinfo{booktitle}{\emph{Proceedings of the Conference on Fairness, Accountability, and Transparency}}. \bibinfo{pages}{359--368}.
\newblock


\bibitem[Murphy et~al\mbox{.}(2025)]%
        {murphy2025sharing}
\bibfield{author}{\bibinfo{person}{Stephanie Murphy}, \bibinfo{person}{Ava Hickey}, \bibinfo{person}{Doireann Peelo~Dennehy}, \bibinfo{person}{Kellie Morrissey}, \bibinfo{person}{John McCarthy}, {and} \bibinfo{person}{Sarah Foley}.} \bibinfo{year}{2025}\natexlab{}.
\newblock \showarticletitle{Sharing, Support-Seeking, and Managing Safety: A Qualitative Study of Online Platform Engagement After Pregnancy Loss}.
\newblock \bibinfo{journal}{\emph{Proceedings of the ACM on Human-Computer Interaction}} \bibinfo{volume}{9}, \bibinfo{number}{2} (\bibinfo{year}{2025}), \bibinfo{pages}{1--26}.
\newblock


\bibitem[Musgrave et~al\mbox{.}(2022)]%
        {musgrave2022experiences}
\bibfield{author}{\bibinfo{person}{Tyler Musgrave}, \bibinfo{person}{Alia Cummings}, {and} \bibinfo{person}{Sarita Schoenebeck}.} \bibinfo{year}{2022}\natexlab{}.
\newblock \showarticletitle{Experiences of Harm, Healing, and Joy among Black Women and Femmes on Social Media}. In \bibinfo{booktitle}{\emph{Proceedings of the 2022 CHI Conference on Human Factors in Computing Systems}}. \bibinfo{pages}{1--17}.
\newblock


\bibitem[O'Brien et~al\mbox{.}(2025)]%
        {o2025online}
\bibfield{author}{\bibinfo{person}{Catherine~RK O'Brien}, \bibinfo{person}{Nuur~Alifah Roslan}, \bibinfo{person}{Steven~J Murdoch}, \bibinfo{person}{Ruba Abu-Salma}, \bibinfo{person}{Douglas Zytko}, {and} \bibinfo{person}{Mark Warner}.} \bibinfo{year}{2025}\natexlab{}.
\newblock \showarticletitle{Online Dating Platform Safeguards and Self-Protection: How Dating Platforms Characterise, Respond to, and Safeguard Against Harms}. In \bibinfo{booktitle}{\emph{Proceedings of the Extended Abstracts of the CHI Conference on Human Factors in Computing Systems}}. \bibinfo{pages}{1--8}.
\newblock


\bibitem[Oguine et~al\mbox{.}(2025)]%
        {oguine2025online}
\bibfield{author}{\bibinfo{person}{Ozioma~Collins Oguine}, \bibinfo{person}{Oghenemaro Anuyah}, \bibinfo{person}{Zainab Agha}, \bibinfo{person}{Iris Melgarez}, \bibinfo{person}{Adriana Alvarado~Garcia}, {and} \bibinfo{person}{Karla Badillo-Urquiola}.} \bibinfo{year}{2025}\natexlab{}.
\newblock \showarticletitle{Online safety for all: Sociocultural insights from a systematic review of youth online safety in the global south}.
\newblock \bibinfo{journal}{\emph{Proceedings of the ACM on Human-Computer Interaction}} \bibinfo{volume}{9}, \bibinfo{number}{7} (\bibinfo{year}{2025}), \bibinfo{pages}{1--30}.
\newblock


\bibitem[Oliver-Smith(1996)]%
        {oliver1996anthropological}
\bibfield{author}{\bibinfo{person}{Anthony Oliver-Smith}.} \bibinfo{year}{1996}\natexlab{}.
\newblock \showarticletitle{Anthropological research on hazards and disasters}.
\newblock \bibinfo{journal}{\emph{Annual review of anthropology}} \bibinfo{volume}{25}, \bibinfo{number}{1} (\bibinfo{year}{1996}), \bibinfo{pages}{303--328}.
\newblock


\bibitem[Pan et~al\mbox{.}(2018)]%
        {pan2018you}
\bibfield{author}{\bibinfo{person}{Wenjing Pan}, \bibinfo{person}{Bo Feng}, {and} \bibinfo{person}{V Skye~Wingate}.} \bibinfo{year}{2018}\natexlab{}.
\newblock \showarticletitle{What you say is what you get: How self-disclosure in support seeking affects language use in support provision in online support forums}.
\newblock \bibinfo{journal}{\emph{Journal of Language and Social Psychology}} \bibinfo{volume}{37}, \bibinfo{number}{1} (\bibinfo{year}{2018}), \bibinfo{pages}{3--27}.
\newblock


\bibitem[Pan et~al\mbox{.}(2020)]%
        {pan2020say}
\bibfield{author}{\bibinfo{person}{Wenjing Pan}, \bibinfo{person}{Bo Feng}, \bibinfo{person}{V~Skye Wingate}, {and} \bibinfo{person}{Siyue Li}.} \bibinfo{year}{2020}\natexlab{}.
\newblock \showarticletitle{What to say when seeking support online: A comparison among different levels of self-disclosure}.
\newblock \bibinfo{journal}{\emph{Frontiers in psychology}}  \bibinfo{volume}{11} (\bibinfo{year}{2020}), \bibinfo{pages}{978}.
\newblock


\bibitem[Pasupathi and Billitteri(2015)]%
        {pasupathi2015being}
\bibfield{author}{\bibinfo{person}{Monisha Pasupathi} {and} \bibinfo{person}{Jacob Billitteri}.} \bibinfo{year}{2015}\natexlab{}.
\newblock \showarticletitle{Being and Becoming through Being Heard: Listener Effects on Stories and Selves}.
\newblock \bibinfo{journal}{\emph{International Journal of Listening}} \bibinfo{volume}{29}, \bibinfo{number}{2} (\bibinfo{year}{2015}), \bibinfo{pages}{67--84}.
\newblock
\href{https://doi.org/10.1080/10904018.2015.1029363}{doi:\nolinkurl{10.1080/10904018.2015.1029363}}


\bibitem[Pasupathi and Hoyt(2009)]%
        {pasupathi2009development}
\bibfield{author}{\bibinfo{person}{Monisha Pasupathi} {and} \bibinfo{person}{Timothy Hoyt}.} \bibinfo{year}{2009}\natexlab{}.
\newblock \showarticletitle{The Development of Narrative Identity in Late Adolescence and Emergent Adulthood: The Continued Importance of Listeners}.
\newblock \bibinfo{journal}{\emph{Developmental Psychology}} \bibinfo{volume}{45}, \bibinfo{number}{2} (\bibinfo{year}{2009}), \bibinfo{pages}{558--574}.
\newblock
\href{https://doi.org/10.1037/a0014431}{doi:\nolinkurl{10.1037/a0014431}}


\bibitem[Pasupathi et~al\mbox{.}(2009)]%
        {pasupathi2009tell}
\bibfield{author}{\bibinfo{person}{Monisha Pasupathi}, \bibinfo{person}{Kate~C. McLean}, {and} \bibinfo{person}{Trisha Weeks}.} \bibinfo{year}{2009}\natexlab{}.
\newblock \showarticletitle{To Tell or Not to Tell: Disclosure and the Narrative Self}.
\newblock \bibinfo{journal}{\emph{Journal of Personality}} \bibinfo{volume}{77}, \bibinfo{number}{1} (\bibinfo{year}{2009}), \bibinfo{pages}{89--124}.
\newblock
\href{https://doi.org/10.1111/j.1467-6494.2008.00539.x}{doi:\nolinkurl{10.1111/j.1467-6494.2008.00539.x}}


\bibitem[Payne et~al\mbox{.}(2025)]%
        {10.1145/3701186}
\bibfield{author}{\bibinfo{person}{Blakeley~H. Payne}, \bibinfo{person}{Jordan Taylor}, \bibinfo{person}{Katta Spiel}, {and} \bibinfo{person}{Casey Fiesler}.} \bibinfo{year}{2025}\natexlab{}.
\newblock \showarticletitle{Building Solidarity Amid Hostility: Experiences of Fat People in Online Communities}.
\newblock \bibinfo{journal}{\emph{Proc. ACM Hum.-Comput. Interact.}} \bibinfo{volume}{9}, \bibinfo{number}{1}, Article \bibinfo{articleno}{GROUP7} (\bibinfo{date}{Jan.} \bibinfo{year}{2025}), \bibinfo{numpages}{27}~pages.
\newblock
\href{https://doi.org/10.1145/3701186}{doi:\nolinkurl{10.1145/3701186}}


\bibitem[Phadke and Mitra(2020)]%
        {phadke2020many}
\bibfield{author}{\bibinfo{person}{Shruti Phadke} {and} \bibinfo{person}{Tanushree Mitra}.} \bibinfo{year}{2020}\natexlab{}.
\newblock \showarticletitle{Many faced hate: A cross platform study of content framing and information sharing by online hate groups}. In \bibinfo{booktitle}{\emph{Proceedings of the 2020 CHI conference on human factors in computing systems}}. \bibinfo{pages}{1--13}.
\newblock


\bibitem[Pinch et~al\mbox{.}(2024)]%
        {pinch2024subtleties}
\bibfield{author}{\bibinfo{person}{Annika Pinch}, \bibinfo{person}{Jeremy Birnholtz}, \bibinfo{person}{Kathryn Macapagal}, \bibinfo{person}{Ashley Kraus}, {and} \bibinfo{person}{David Moskowitz}.} \bibinfo{year}{2024}\natexlab{}.
\newblock \showarticletitle{The subtleties of self-presentation: A study of sensitive disclosure among sexual minority adolescents}.
\newblock \bibinfo{journal}{\emph{Proceedings of the ACM on Human-Computer Interaction}} \bibinfo{volume}{8}, \bibinfo{number}{CSCW1} (\bibinfo{year}{2024}), \bibinfo{pages}{1--27}.
\newblock


\bibitem[Pretorius et~al\mbox{.}(2020)]%
        {pretorius2020searching}
\bibfield{author}{\bibinfo{person}{Claudette Pretorius}, \bibinfo{person}{Darragh McCashin}, \bibinfo{person}{Naoise Kavanagh}, {and} \bibinfo{person}{David Coyle}.} \bibinfo{year}{2020}\natexlab{}.
\newblock \showarticletitle{Searching for mental health: a mixed-methods study of young people's online help-seeking}. In \bibinfo{booktitle}{\emph{Proceedings of the 2020 CHI Conference on Human Factors in Computing Systems}}. \bibinfo{pages}{1--13}.
\newblock


\bibitem[Progga et~al\mbox{.}(2023)]%
        {progga2023understanding}
\bibfield{author}{\bibinfo{person}{Farhat~Tasnim Progga}, \bibinfo{person}{Avanthika Senthil~Kumar}, {and} \bibinfo{person}{Sabirat Rubya}.} \bibinfo{year}{2023}\natexlab{}.
\newblock \showarticletitle{Understanding the Online Social Support Dynamics for Postpartum Depression}. In \bibinfo{booktitle}{\emph{Proceedings of the 2023 CHI Conference on Human Factors in Computing Systems}}. \bibinfo{pages}{1--17}.
\newblock


\bibitem[Quinn and Gutt(2025)]%
        {quinn2025heterogeneous}
\bibfield{author}{\bibinfo{person}{Martin Quinn} {and} \bibinfo{person}{Dominik Gutt}.} \bibinfo{year}{2025}\natexlab{}.
\newblock \showarticletitle{Heterogeneous effects of generative artificial intelligence (GenAI) on knowledge seeking in online communities}.
\newblock \bibinfo{journal}{\emph{Journal of Management Information Systems}} \bibinfo{volume}{42}, \bibinfo{number}{2} (\bibinfo{year}{2025}), \bibinfo{pages}{370--399}.
\newblock


\bibitem[Rajadesingan et~al\mbox{.}(2020)]%
        {rajadesingan2020quick}
\bibfield{author}{\bibinfo{person}{Ashwin Rajadesingan}, \bibinfo{person}{Paul Resnick}, {and} \bibinfo{person}{Ceren Budak}.} \bibinfo{year}{2020}\natexlab{}.
\newblock \showarticletitle{Quick, community-specific learning: How distinctive toxicity norms are maintained in political subreddits}. In \bibinfo{booktitle}{\emph{Proceedings of the International AAAI Conference on Web and Social Media}}, Vol.~\bibinfo{volume}{14}. \bibinfo{pages}{557--568}.
\newblock


\bibitem[Rho et~al\mbox{.}(2017)]%
        {rho2017class}
\bibfield{author}{\bibinfo{person}{Eugenia Ha~Rim Rho}, \bibinfo{person}{Oliver~L Haimson}, \bibinfo{person}{Nazanin Andalibi}, \bibinfo{person}{Melissa Mazmanian}, {and} \bibinfo{person}{Gillian~R Hayes}.} \bibinfo{year}{2017}\natexlab{}.
\newblock \showarticletitle{Class confessions: Restorative properties in online experiences of socioeconomic stigma}. In \bibinfo{booktitle}{\emph{Proceedings of the 2017 CHI Conference on Human Factors in Computing Systems}}. \bibinfo{pages}{3377--3389}.
\newblock


\bibitem[Ricoeur(1980)]%
        {ricoeur1980narrative}
\bibfield{author}{\bibinfo{person}{Paul Ricoeur}.} \bibinfo{year}{1980}\natexlab{}.
\newblock \showarticletitle{Narrative Time}.
\newblock \bibinfo{journal}{\emph{Critical Inquiry}} \bibinfo{volume}{7}, \bibinfo{number}{1} (\bibinfo{year}{1980}), \bibinfo{pages}{169--190}.
\newblock
\href{https://doi.org/10.1086/448093}{doi:\nolinkurl{10.1086/448093}}


\bibitem[Ricoeur(1991)]%
        {ricoeur1991human}
\bibfield{author}{\bibinfo{person}{Paul Ricoeur}.} \bibinfo{year}{1991}\natexlab{}.
\newblock \showarticletitle{The Human Experience of Time and Narrative}.
\newblock In \bibinfo{booktitle}{\emph{A Ricoeur Reader: Reflection and Imagination}}, \bibfield{editor}{\bibinfo{person}{Mario~J. Vald{\'e}s}} (Ed.). \bibinfo{publisher}{University of Toronto Press}, \bibinfo{address}{Toronto}, \bibinfo{pages}{99--116}.
\newblock


\bibitem[Rodriguez and Cohen(1998)]%
        {rodriguez1998social}
\bibfield{author}{\bibinfo{person}{Mario~S Rodriguez} {and} \bibinfo{person}{Sheldon Cohen}.} \bibinfo{year}{1998}\natexlab{}.
\newblock \showarticletitle{Social support}.
\newblock \bibinfo{journal}{\emph{Encyclopedia of mental health}} \bibinfo{volume}{3}, \bibinfo{number}{2} (\bibinfo{year}{1998}), \bibinfo{pages}{535--544}.
\newblock


\bibitem[Rubin et~al\mbox{.}(2020)]%
        {rubin2020fragile}
\bibfield{author}{\bibinfo{person}{Jennifer~D Rubin}, \bibinfo{person}{Lindsay Blackwell}, {and} \bibinfo{person}{Terri~D Conley}.} \bibinfo{year}{2020}\natexlab{}.
\newblock \showarticletitle{Fragile masculinity: Men, gender, and online harassment}. In \bibinfo{booktitle}{\emph{Proceedings of the 2020 CHI conference on human factors in computing systems}}. \bibinfo{pages}{1--14}.
\newblock


\bibitem[Ryu and Kang(2025)]%
        {ryu2025exploring}
\bibfield{author}{\bibinfo{person}{Hyeyoung Ryu} {and} \bibinfo{person}{Sungha Kang}.} \bibinfo{year}{2025}\natexlab{}.
\newblock \showarticletitle{Exploring Design Tensions in Countering Microaggressions in Online Communities for Stigmatized Health Support}. In \bibinfo{booktitle}{\emph{Proceedings of the Extended Abstracts of the CHI Conference on Human Factors in Computing Systems}}. \bibinfo{pages}{1--9}.
\newblock


\bibitem[Saha et~al\mbox{.}(2020)]%
        {saha2020understanding}
\bibfield{author}{\bibinfo{person}{Koustuv Saha}, \bibinfo{person}{Sindhu~Kiranmai Ernala}, \bibinfo{person}{Sarmistha Dutta}, \bibinfo{person}{Eva Sharma}, {and} \bibinfo{person}{Munmun De~Choudhury}.} \bibinfo{year}{2020}\natexlab{}.
\newblock \showarticletitle{Understanding moderation in online mental health communities}. In \bibinfo{booktitle}{\emph{International Conference on Human-Computer Interaction}}. Springer, \bibinfo{pages}{87--107}.
\newblock


\bibitem[Saldaña(2013)]%
        {saldana_coding_2013}
\bibfield{author}{\bibinfo{person}{Johnny Saldaña}.} \bibinfo{year}{2013}\natexlab{}.
\newblock \bibinfo{booktitle}{\emph{The coding manual for qualitative researchers} (\bibinfo{edition}{2nd ed} ed.)}.
\newblock \bibinfo{publisher}{SAGE}, \bibinfo{address}{Los Angeles}.
\newblock
\showISBNx{978-1-4462-4736-5 978-1-4462-4737-2}
\newblock
\shownote{OCLC: ocn796279115}.


\bibitem[Scheuerman et~al\mbox{.}(2021)]%
        {scheuerman2021framework}
\bibfield{author}{\bibinfo{person}{Morgan~Klaus Scheuerman}, \bibinfo{person}{Jialun~Aaron Jiang}, \bibinfo{person}{Casey Fiesler}, {and} \bibinfo{person}{Jed~R Brubaker}.} \bibinfo{year}{2021}\natexlab{}.
\newblock \showarticletitle{A framework of severity for harmful content online}.
\newblock \bibinfo{journal}{\emph{Proceedings of the ACM on Human-Computer Interaction}} \bibinfo{volume}{5}, \bibinfo{number}{CSCW2} (\bibinfo{year}{2021}), \bibinfo{pages}{1--33}.
\newblock


\bibitem[Scott et~al\mbox{.}(2023)]%
        {scott2023trauma}
\bibfield{author}{\bibinfo{person}{Carol~F Scott}, \bibinfo{person}{Gabriela Marcu}, \bibinfo{person}{Riana~Elyse Anderson}, \bibinfo{person}{Mark~W Newman}, {and} \bibinfo{person}{Sarita Schoenebeck}.} \bibinfo{year}{2023}\natexlab{}.
\newblock \showarticletitle{Trauma-Informed Social Media: Towards Solutions for Reducing and Healing Online Harm}. In \bibinfo{booktitle}{\emph{Proceedings of the 2023 CHI Conference on Human Factors in Computing Systems}}. \bibinfo{pages}{1--20}.
\newblock


\bibitem[Sha et~al\mbox{.}(2025)]%
        {sha2025meeting}
\bibfield{author}{\bibinfo{person}{Yongjie Sha}, \bibinfo{person}{Alexis Roth}, \bibinfo{person}{Stephen Lankenau}, \bibinfo{person}{David~G Schwartz}, {and} \bibinfo{person}{Gabriela Marcu}.} \bibinfo{year}{2025}\natexlab{}.
\newblock \showarticletitle{Meeting People Where They Are: Building Community-Centered Care with Smartphone-Facilitated Response to Overdoses}.
\newblock \bibinfo{journal}{\emph{Proceedings of the ACM on Human-Computer Interaction}} \bibinfo{volume}{9}, \bibinfo{number}{1} (\bibinfo{year}{2025}), \bibinfo{pages}{1--23}.
\newblock


\bibitem[Sharma and De~Choudhury(2018)]%
        {sharma2018mental}
\bibfield{author}{\bibinfo{person}{Eva Sharma} {and} \bibinfo{person}{Munmun De~Choudhury}.} \bibinfo{year}{2018}\natexlab{}.
\newblock \showarticletitle{Mental health support and its relationship to linguistic accommodation in online communities}. In \bibinfo{booktitle}{\emph{Proceedings of the 2018 CHI conference on human factors in computing systems}}. \bibinfo{pages}{1--13}.
\newblock


\bibitem[Shelby et~al\mbox{.}(2023)]%
        {shelby2023sociotechnical}
\bibfield{author}{\bibinfo{person}{Renee Shelby}, \bibinfo{person}{Shalaleh Rismani}, \bibinfo{person}{Kathryn Henne}, \bibinfo{person}{AJung Moon}, \bibinfo{person}{Negar Rostamzadeh}, \bibinfo{person}{Paul Nicholas}, \bibinfo{person}{N'Mah Yilla-Akbari}, \bibinfo{person}{Jess Gallegos}, \bibinfo{person}{Andrew Smart}, \bibinfo{person}{Emilio Garcia}, {et~al\mbox{.}}} \bibinfo{year}{2023}\natexlab{}.
\newblock \showarticletitle{Sociotechnical harms of algorithmic systems: Scoping a taxonomy for harm reduction}. In \bibinfo{booktitle}{\emph{Proceedings of the 2023 AAAI/ACM Conference on AI, Ethics, and Society}}. \bibinfo{pages}{723--741}.
\newblock


\bibitem[Shrider and Creamer(2023)]%
        {shrider2023poverty}
\bibfield{author}{\bibinfo{person}{Emily~A Shrider} {and} \bibinfo{person}{John Creamer}.} \bibinfo{year}{2023}\natexlab{}.
\newblock \showarticletitle{Poverty in the United States: 2022}.
\newblock \bibinfo{journal}{\emph{US Census Bureau}} (\bibinfo{year}{2023}).
\newblock


\bibitem[Simpson and Semaan(2021)]%
        {simpson2021you}
\bibfield{author}{\bibinfo{person}{Ellen Simpson} {and} \bibinfo{person}{Bryan Semaan}.} \bibinfo{year}{2021}\natexlab{}.
\newblock \showarticletitle{For you, or for" you"? Everyday LGBTQ+ encounters with TikTok}.
\newblock \bibinfo{journal}{\emph{Proceedings of the ACM on human-computer interaction}} \bibinfo{volume}{4}, \bibinfo{number}{CSCW3} (\bibinfo{year}{2021}), \bibinfo{pages}{1--34}.
\newblock


\bibitem[Singer et~al\mbox{.}(2014)]%
        {singer2014evolution}
\bibfield{author}{\bibinfo{person}{Philipp Singer}, \bibinfo{person}{Fabian Fl{\"o}ck}, \bibinfo{person}{Clemens Meinhart}, \bibinfo{person}{Elias Zeitfogel}, {and} \bibinfo{person}{Markus Strohmaier}.} \bibinfo{year}{2014}\natexlab{}.
\newblock \showarticletitle{Evolution of reddit: from the front page of the internet to a self-referential community?}. In \bibinfo{booktitle}{\emph{Proceedings of the 23rd international conference on world wide web}}. \bibinfo{pages}{517--522}.
\newblock


\bibitem[Singh et~al\mbox{.}(2017)]%
        {singh2017they}
\bibfield{author}{\bibinfo{person}{Vivek~K Singh}, \bibinfo{person}{Marie~L Radford}, \bibinfo{person}{Qianjia Huang}, {and} \bibinfo{person}{Susan Furrer}.} \bibinfo{year}{2017}\natexlab{}.
\newblock \showarticletitle{" They basically like destroyed the school one day" On Newer App Features and Cyberbullying in Schools}. In \bibinfo{booktitle}{\emph{Proceedings of the 2017 ACM Conference on Computer Supported Cooperative Work and Social Computing}}. \bibinfo{pages}{1210--1216}.
\newblock


\bibitem[Skeels et~al\mbox{.}(2010)]%
        {skeels2010catalyzing}
\bibfield{author}{\bibinfo{person}{Meredith~M Skeels}, \bibinfo{person}{Kenton~T Unruh}, \bibinfo{person}{Christopher Powell}, {and} \bibinfo{person}{Wanda Pratt}.} \bibinfo{year}{2010}\natexlab{}.
\newblock \showarticletitle{Catalyzing social support for breast cancer patients}. In \bibinfo{booktitle}{\emph{Proceedings of the SIGCHI Conference on Human Factors in Computing Systems}}. \bibinfo{pages}{173--182}.
\newblock


\bibitem[Sultana et~al\mbox{.}(2022)]%
        {sultana2022imagined}
\bibfield{author}{\bibinfo{person}{Sharifa Sultana}, \bibinfo{person}{Pratyasha Saha}, \bibinfo{person}{Shaid Hasan}, \bibinfo{person}{SM~Raihanul Alam}, \bibinfo{person}{Rokeya Akter}, \bibinfo{person}{Md~Mirajul Islam}, \bibinfo{person}{Raihan~Islam Arnob}, \bibinfo{person}{AKM~Najmul Islam}, \bibinfo{person}{Mahdi~Nasrullah Al-Ameen}, {and} \bibinfo{person}{Syed~Ishtiaque Ahmed}.} \bibinfo{year}{2022}\natexlab{}.
\newblock \showarticletitle{Imagined online communities: Communionship, sovereignty, and inclusiveness in Facebook Groups}.
\newblock \bibinfo{journal}{\emph{Proceedings of the ACM on Human-Computer Interaction}} \bibinfo{volume}{6}, \bibinfo{number}{CSCW2} (\bibinfo{year}{2022}), \bibinfo{pages}{1--29}.
\newblock


\bibitem[Suresh and Guttag(2021)]%
        {suresh2021framework}
\bibfield{author}{\bibinfo{person}{Harini Suresh} {and} \bibinfo{person}{John Guttag}.} \bibinfo{year}{2021}\natexlab{}.
\newblock \showarticletitle{A framework for understanding sources of harm throughout the machine learning life cycle}. In \bibinfo{booktitle}{\emph{Proceedings of the 1st ACM Conference on Equity and Access in Algorithms, Mechanisms, and Optimization}}. \bibinfo{pages}{1--9}.
\newblock


\bibitem[Take et~al\mbox{.}(2024)]%
        {take2024stoking}
\bibfield{author}{\bibinfo{person}{Kejsi Take}, \bibinfo{person}{Victoria Zhong}, \bibinfo{person}{Chris Geeng}, \bibinfo{person}{Emmi Bevensee}, \bibinfo{person}{Damon McCoy}, {and} \bibinfo{person}{Rachel Greenstadt}.} \bibinfo{year}{2024}\natexlab{}.
\newblock \showarticletitle{Stoking the Flames: Understanding Escalation in an Online Harassment Community}.
\newblock \bibinfo{journal}{\emph{Proceedings of the ACM on Human-Computer Interaction}} \bibinfo{volume}{8}, \bibinfo{number}{CSCW1} (\bibinfo{year}{2024}), \bibinfo{pages}{1--23}.
\newblock


\bibitem[Tang et~al\mbox{.}(2025)]%
        {10.1145/3710935}
\bibfield{author}{\bibinfo{person}{Xinru Tang}, \bibinfo{person}{Gabriel Lima}, \bibinfo{person}{Li Jiang}, \bibinfo{person}{Lucy Simko}, {and} \bibinfo{person}{Yixin Zou}.} \bibinfo{year}{2025}\natexlab{}.
\newblock \showarticletitle{Beyond "Vulnerable Populations": A Unified Understanding of Vulnerability From A Socio-Ecological Perspective}.
\newblock \bibinfo{journal}{\emph{Proc. ACM Hum.-Comput. Interact.}} \bibinfo{volume}{9}, \bibinfo{number}{2}, Article \bibinfo{articleno}{CSCW037} (\bibinfo{date}{May} \bibinfo{year}{2025}), \bibinfo{numpages}{30}~pages.
\newblock
\href{https://doi.org/10.1145/3710935}{doi:\nolinkurl{10.1145/3710935}}


\bibitem[Tanni et~al\mbox{.}(2024)]%
        {tanni2024examining}
\bibfield{author}{\bibinfo{person}{Tangila~Islam Tanni}, \bibinfo{person}{Mamtaj Akter}, \bibinfo{person}{Joshua Anderson}, \bibinfo{person}{Mary~Jean Amon}, {and} \bibinfo{person}{Pamela~J Wisniewski}.} \bibinfo{year}{2024}\natexlab{}.
\newblock \showarticletitle{Examining the unique online risk experiences and mental health outcomes of lgbtq+ versus heterosexual youth}. In \bibinfo{booktitle}{\emph{Proceedings of the 2024 CHI Conference on Human Factors in Computing Systems}}. \bibinfo{pages}{1--21}.
\newblock


\bibitem[Taylor et~al\mbox{.}(2024)]%
        {taylor2024cruising}
\bibfield{author}{\bibinfo{person}{Jordan Taylor}, \bibinfo{person}{Ellen Simpson}, \bibinfo{person}{Anh-Ton Tran}, \bibinfo{person}{Jed~R Brubaker}, \bibinfo{person}{Sarah~E Fox}, {and} \bibinfo{person}{Haiyi Zhu}.} \bibinfo{year}{2024}\natexlab{}.
\newblock \showarticletitle{Cruising Queer HCI on the DL: A literature review of LGBTQ+ people in HCI}. In \bibinfo{booktitle}{\emph{Proceedings of the 2024 CHI Conference on Human Factors in Computing Systems}}. \bibinfo{pages}{1--21}.
\newblock


\bibitem[Thebault-Spieker et~al\mbox{.}(2023)]%
        {thebault2023diverse}
\bibfield{author}{\bibinfo{person}{Jacob Thebault-Spieker}, \bibinfo{person}{Sukrit Venkatagiri}, \bibinfo{person}{Naomi Mine}, {and} \bibinfo{person}{Kurt Luther}.} \bibinfo{year}{2023}\natexlab{}.
\newblock \showarticletitle{Diverse perspectives can mitigate political bias in crowdsourced content moderation}. In \bibinfo{booktitle}{\emph{Proceedings of the 2023 ACM Conference on Fairness, Accountability, and Transparency}}. \bibinfo{pages}{1280--1291}.
\newblock


\bibitem[Thoits(1995)]%
        {thoits1995stress}
\bibfield{author}{\bibinfo{person}{Peggy~A Thoits}.} \bibinfo{year}{1995}\natexlab{}.
\newblock \showarticletitle{Stress, coping, and social support processes: Where are we? What next?}
\newblock \bibinfo{journal}{\emph{Journal of health and social behavior}} (\bibinfo{year}{1995}), \bibinfo{pages}{53--79}.
\newblock


\bibitem[Thomas et~al\mbox{.}(2022)]%
        {thomas2022s}
\bibfield{author}{\bibinfo{person}{Kurt Thomas}, \bibinfo{person}{Patrick~Gage Kelley}, \bibinfo{person}{Sunny Consolvo}, \bibinfo{person}{Patrawat Samermit}, {and} \bibinfo{person}{Elie Bursztein}.} \bibinfo{year}{2022}\natexlab{}.
\newblock \showarticletitle{“It’s common and a part of being a content creator”: Understanding How Creators Experience and Cope with Hate and Harassment Online}. In \bibinfo{booktitle}{\emph{Proceedings of the 2022 CHI Conference on Human Factors in Computing Systems}}. \bibinfo{pages}{1--15}.
\newblock


\bibitem[Tomprou et~al\mbox{.}(2019)]%
        {tomprou2019career}
\bibfield{author}{\bibinfo{person}{Maria Tomprou}, \bibinfo{person}{Laura Dabbish}, \bibinfo{person}{Robert~E Kraut}, {and} \bibinfo{person}{Fannie Liu}.} \bibinfo{year}{2019}\natexlab{}.
\newblock \showarticletitle{Career mentoring in online communities: Seeking and receiving advice from an online community}. In \bibinfo{booktitle}{\emph{Proceedings of the 2019 CHI Conference on Human Factors in Computing Systems}}. \bibinfo{pages}{1--12}.
\newblock


\bibitem[Uttarapong et~al\mbox{.}(2024)]%
        {uttarapong2024moderatorhub}
\bibfield{author}{\bibinfo{person}{Jirassaya Uttarapong}, \bibinfo{person}{Yihan Wang}, {and} \bibinfo{person}{Donghee~Yvette Wohn}.} \bibinfo{year}{2024}\natexlab{}.
\newblock \showarticletitle{ModeratorHub: A Knowledge Sharing and Relationship Building Platform for Moderators}. In \bibinfo{booktitle}{\emph{Extended Abstracts of the CHI Conference on Human Factors in Computing Systems}}. \bibinfo{pages}{1--6}.
\newblock


\bibitem[Van~Royen et~al\mbox{.}(2017)]%
        {van2017thinking}
\bibfield{author}{\bibinfo{person}{Kathleen Van~Royen}, \bibinfo{person}{Karolien Poels}, \bibinfo{person}{Heidi Vandebosch}, {and} \bibinfo{person}{Philippe Adam}.} \bibinfo{year}{2017}\natexlab{}.
\newblock \showarticletitle{“Thinking before posting?” Reducing cyber harassment on social networking sites through a reflective message}.
\newblock \bibinfo{journal}{\emph{Computers in human behavior}}  \bibinfo{volume}{66} (\bibinfo{year}{2017}), \bibinfo{pages}{345--352}.
\newblock


\bibitem[Vitak et~al\mbox{.}(2017)]%
        {vitak2017identifying}
\bibfield{author}{\bibinfo{person}{Jessica Vitak}, \bibinfo{person}{Kalyani Chadha}, \bibinfo{person}{Linda Steiner}, {and} \bibinfo{person}{Zahra Ashktorab}.} \bibinfo{year}{2017}\natexlab{}.
\newblock \showarticletitle{Identifying women's experiences with and strategies for mitigating negative effects of online harassment}. In \bibinfo{booktitle}{\emph{Proceedings of the 2017 ACM Conference on Computer Supported Cooperative Work and Social Computing}}. \bibinfo{pages}{1231--1245}.
\newblock


\bibitem[Wakefield et~al\mbox{.}(2013)]%
        {wakefield2013meta}
\bibfield{author}{\bibinfo{person}{Juliet~RH Wakefield}, \bibinfo{person}{Nick Hopkins}, {and} \bibinfo{person}{Ronni~Michelle Greenwood}.} \bibinfo{year}{2013}\natexlab{}.
\newblock \showarticletitle{Meta-stereotypes, social image and help seeking: Dependency-related meta-stereotypes reduce help-seeking behaviour}.
\newblock \bibinfo{journal}{\emph{Journal of community \& applied social psychology}} \bibinfo{volume}{23}, \bibinfo{number}{5} (\bibinfo{year}{2013}), \bibinfo{pages}{363--372}.
\newblock


\bibitem[Wei et~al\mbox{.}(2023)]%
        {wei2023there}
\bibfield{author}{\bibinfo{person}{Miranda Wei}, \bibinfo{person}{Sunny Consolvo}, \bibinfo{person}{Patrick~Gage Kelley}, \bibinfo{person}{Tadayoshi Kohno}, \bibinfo{person}{Franziska Roesner}, {and} \bibinfo{person}{Kurt Thomas}.} \bibinfo{year}{2023}\natexlab{}.
\newblock \showarticletitle{“There’s so much responsibility on users right now:” Expert Advice for Staying Safer From Hate and Harassment}. In \bibinfo{booktitle}{\emph{Proceedings of the 2023 CHI Conference on Human Factors in Computing Systems}}. \bibinfo{pages}{1--17}.
\newblock


\bibitem[Whittaker and Kowalski(2015)]%
        {whittaker2015cyberbullying}
\bibfield{author}{\bibinfo{person}{Elizabeth Whittaker} {and} \bibinfo{person}{Robin~M Kowalski}.} \bibinfo{year}{2015}\natexlab{}.
\newblock \showarticletitle{Cyberbullying via social media}.
\newblock \bibinfo{journal}{\emph{Journal of school violence}} \bibinfo{volume}{14}, \bibinfo{number}{1} (\bibinfo{year}{2015}), \bibinfo{pages}{11--29}.
\newblock


\bibitem[Williams et~al\mbox{.}(2025)]%
        {williams2025can}
\bibfield{author}{\bibinfo{person}{Gianna Williams}, \bibinfo{person}{Natalie Chen}, \bibinfo{person}{Michael~Ann DeVito}, {and} \bibinfo{person}{Alexandra To}.} \bibinfo{year}{2025}\natexlab{}.
\newblock \showarticletitle{Why Can't Black Women Just Be?: Black Femme Content Creators Navigating Algorithmic Monoliths}. In \bibinfo{booktitle}{\emph{Proceedings of the 2025 CHI Conference on Human Factors in Computing Systems}}. \bibinfo{pages}{1--14}.
\newblock


\bibitem[Williams(2007)]%
        {williams2007ostracism}
\bibfield{author}{\bibinfo{person}{Kipling~D Williams}.} \bibinfo{year}{2007}\natexlab{}.
\newblock \showarticletitle{Ostracism}.
\newblock \bibinfo{journal}{\emph{Annu. Rev. Psychol.}} \bibinfo{volume}{58}, \bibinfo{number}{1} (\bibinfo{year}{2007}), \bibinfo{pages}{425--452}.
\newblock


\bibitem[Williams and Clarke(2019)]%
        {williams2019desensitization}
\bibfield{author}{\bibinfo{person}{Stephanie~N Williams} {and} \bibinfo{person}{Annette~V Clarke}.} \bibinfo{year}{2019}\natexlab{}.
\newblock \showarticletitle{How the desensitization of police violence, stereotyped language, and racial bias impact black communities}.
\newblock \bibinfo{journal}{\emph{Psychology and Cognitive Sciences--Open Journal}} \bibinfo{volume}{5}, \bibinfo{number}{2} (\bibinfo{year}{2019}), \bibinfo{pages}{62--67}.
\newblock


\bibitem[Wingate et~al\mbox{.}(2020)]%
        {wingate2020influence}
\bibfield{author}{\bibinfo{person}{V~Skye Wingate}, \bibinfo{person}{Bo Feng}, \bibinfo{person}{Chelsea Kim}, \bibinfo{person}{Wenjing Pan}, {and} \bibinfo{person}{JooYoung Jang}.} \bibinfo{year}{2020}\natexlab{}.
\newblock \showarticletitle{The influence of self-disclosure in online support seeking on quality of received advice}.
\newblock \bibinfo{journal}{\emph{Journal of Language and Social Psychology}} \bibinfo{volume}{39}, \bibinfo{number}{3} (\bibinfo{year}{2020}), \bibinfo{pages}{397--413}.
\newblock


\bibitem[Wright et~al\mbox{.}(2003)]%
        {wright2003health}
\bibfield{author}{\bibinfo{person}{Kevin~B Wright}, \bibinfo{person}{Sally~B Bell}, \bibinfo{person}{Kevin~B Wright}, {and} \bibinfo{person}{Sally~B Bell}.} \bibinfo{year}{2003}\natexlab{}.
\newblock \showarticletitle{Health-related support groups on the Internet: Linking empirical findings to social support and computer-mediated communication theory}.
\newblock \bibinfo{journal}{\emph{Journal of health psychology}} \bibinfo{volume}{8}, \bibinfo{number}{1} (\bibinfo{year}{2003}), \bibinfo{pages}{39--54}.
\newblock


\bibitem[Xiao et~al\mbox{.}(2025)]%
        {xiao2025snugglesense}
\bibfield{author}{\bibinfo{person}{Sijia Xiao}, \bibinfo{person}{Haodi Zou}, \bibinfo{person}{Amy Mathews}, \bibinfo{person}{Jingshu Rui}, \bibinfo{person}{Coye Cheshire}, {and} \bibinfo{person}{Niloufar Salehi}.} \bibinfo{year}{2025}\natexlab{}.
\newblock \showarticletitle{SnuggleSense: Empowering Online Harm Survivors Through a Structured Sensemaking Process}.
\newblock \bibinfo{journal}{\emph{Proceedings of the ACM on Human-Computer Interaction}} \bibinfo{volume}{9}, \bibinfo{number}{7} (\bibinfo{year}{2025}), \bibinfo{pages}{1--27}.
\newblock


\bibitem[Xu and Zhang(2025)]%
        {10.1145/3701210}
\bibfield{author}{\bibinfo{person}{Hongdi Xu} {and} \bibinfo{person}{Pengyi Zhang}.} \bibinfo{year}{2025}\natexlab{}.
\newblock \showarticletitle{The Influence of Anonymity and Social Ties on Personal Experience Sharing: A Comprehensive Mixed-Methods Study}.
\newblock \bibinfo{journal}{\emph{Proc. ACM Hum.-Comput. Interact.}} \bibinfo{volume}{9}, \bibinfo{number}{1}, Article \bibinfo{articleno}{GROUP31} (\bibinfo{date}{Jan.} \bibinfo{year}{2025}), \bibinfo{numpages}{22}~pages.
\newblock
\href{https://doi.org/10.1145/3701210}{doi:\nolinkurl{10.1145/3701210}}


\bibitem[Yang et~al\mbox{.}(2019)]%
        {yang2019channel}
\bibfield{author}{\bibinfo{person}{Diyi Yang}, \bibinfo{person}{Zheng Yao}, \bibinfo{person}{Joseph Seering}, {and} \bibinfo{person}{Robert Kraut}.} \bibinfo{year}{2019}\natexlab{}.
\newblock \showarticletitle{The channel matters: Self-disclosure, reciprocity and social support in online cancer support groups}. In \bibinfo{booktitle}{\emph{Proceedings of the 2019 chi conference on human factors in computing systems}}. \bibinfo{pages}{1--15}.
\newblock


\bibitem[Yang et~al\mbox{.}(2025)]%
        {yang2025m}
\bibfield{author}{\bibinfo{person}{Yukun Yang}, \bibinfo{person}{Pratiti Tagore}, {and} \bibinfo{person}{Brooke Foucault~Welles}.} \bibinfo{year}{2025}\natexlab{}.
\newblock \showarticletitle{'I'm so sick and tired of this': Understanding the Roles and Scenarios in Self-Disclosure of Everyday Experiences of Racism on Social Media}.
\newblock \bibinfo{journal}{\emph{Proceedings of the ACM on Human-Computer Interaction}} \bibinfo{volume}{9}, \bibinfo{number}{7} (\bibinfo{year}{2025}), \bibinfo{pages}{1--35}.
\newblock


\bibitem[Yin et~al\mbox{.}(2025)]%
        {yin2025understanding}
\bibfield{author}{\bibinfo{person}{Jiayu~Yuki Yin}, \bibinfo{person}{Novia Wong}, {and} \bibinfo{person}{Madhu Reddy}.} \bibinfo{year}{2025}\natexlab{}.
\newblock \showarticletitle{Understanding User Experience of Support-Seeking on Reddit During Stressful Times}.
\newblock \bibinfo{journal}{\emph{Proceedings of the ACM on Human-Computer Interaction}} \bibinfo{volume}{9}, \bibinfo{number}{7} (\bibinfo{year}{2025}), \bibinfo{pages}{1--25}.
\newblock


\bibitem[Young and Miller(2019)]%
        {young2019girl}
\bibfield{author}{\bibinfo{person}{Alyson~L Young} {and} \bibinfo{person}{Andrew~D Miller}.} \bibinfo{year}{2019}\natexlab{}.
\newblock \showarticletitle{" This Girl is on Fire" Sensemaking in an Online Health Community for Vulvodynia}. In \bibinfo{booktitle}{\emph{Proceedings of the 2019 CHI Conference on Human Factors in Computing Systems}}. \bibinfo{pages}{1--13}.
\newblock


\bibitem[Zhai and Du(2022)]%
        {zhai2022disparities}
\bibfield{author}{\bibinfo{person}{Yusen Zhai} {and} \bibinfo{person}{Xue Du}.} \bibinfo{year}{2022}\natexlab{}.
\newblock \showarticletitle{Disparities and intersectionality in social support networks: addressing social inequalities during the COVID-19 pandemic and beyond}.
\newblock \bibinfo{journal}{\emph{Humanities and Social Sciences Communications}} \bibinfo{volume}{9}, \bibinfo{number}{1} (\bibinfo{year}{2022}), \bibinfo{pages}{1--5}.
\newblock


\bibitem[Zhang(2017)]%
        {zhang2017stress}
\bibfield{author}{\bibinfo{person}{Renwen Zhang}.} \bibinfo{year}{2017}\natexlab{}.
\newblock \showarticletitle{The stress-buffering effect of self-disclosure on Facebook: An examination of stressful life events, social support, and mental health among college students}.
\newblock \bibinfo{journal}{\emph{Computers in Human Behavior}}  \bibinfo{volume}{75} (\bibinfo{year}{2017}), \bibinfo{pages}{527--537}.
\newblock


\bibitem[Zhang et~al\mbox{.}(2025)]%
        {zhang2025fewer}
\bibfield{author}{\bibinfo{person}{Xinyi Zhang}, \bibinfo{person}{Muskan Gupta}, \bibinfo{person}{Emily Altland}, {and} \bibinfo{person}{Sang~Won Lee}.} \bibinfo{year}{2025}\natexlab{}.
\newblock \showarticletitle{'Fewer Views If They Have TW.': Understanding Users' Perceptions of Trigger Warning and Content Warning on Social Media Platforms in the US}.
\newblock \bibinfo{journal}{\emph{Proceedings of the ACM on Human-Computer Interaction}} \bibinfo{volume}{9}, \bibinfo{number}{7} (\bibinfo{year}{2025}), \bibinfo{pages}{1--34}.
\newblock


\bibitem[Zhao et~al\mbox{.}(2013)]%
        {zhao2013many}
\bibfield{author}{\bibinfo{person}{Xuan Zhao}, \bibinfo{person}{Niloufar Salehi}, \bibinfo{person}{Sasha Naranjit}, \bibinfo{person}{Sara Alwaalan}, \bibinfo{person}{Stephen Voida}, {and} \bibinfo{person}{Dan Cosley}.} \bibinfo{year}{2013}\natexlab{}.
\newblock \showarticletitle{The many faces of Facebook: Experiencing social media as performance, exhibition, and personal archive}. In \bibinfo{booktitle}{\emph{Proceedings of the SIGCHI conference on human factors in computing systems}}. \bibinfo{pages}{1--10}.
\newblock


\bibitem[Zheng et~al\mbox{.}(2024)]%
        {zheng2024s}
\bibfield{author}{\bibinfo{person}{Wenqi Zheng}, \bibinfo{person}{Emma Walquist}, \bibinfo{person}{Isha Datey}, \bibinfo{person}{Xiangyu Zhou}, \bibinfo{person}{Kelly Berishaj}, \bibinfo{person}{Melissa Mcdonald}, \bibinfo{person}{Michele Parkhill}, \bibinfo{person}{Dongxiao Zhu}, {and} \bibinfo{person}{Douglas Zytko}.} \bibinfo{year}{2024}\natexlab{}.
\newblock \showarticletitle{“It’s Not What We Were Trying to Get At, but I Think Maybe It Should Be”: Learning How to Do Trauma-Informed Design with a Data Donation Platform for Online Dating Sexual Violence}. In \bibinfo{booktitle}{\emph{Proceedings of the 2024 CHI Conference on Human Factors in Computing Systems}}. \bibinfo{pages}{1--15}.
\newblock


\end{thebibliography}

\appendix
\section{Interview Protocol}
 \subsection{General experience using Reddit}
\begin{itemize}
\item How long have you been using Reddit?

\item Could you describe what you usually do on the platform?

\item How do you interact with other users? How frequently? And why?
How much time do you spend on Reddit?

\item Which subreddits do you visit frequently? For what purpose?

\item What do you think of the community? 
\end{itemize}
  
\subsection{Online support seeking}
\begin{itemize}
\item Which subreddit(s) have you sought support in? What kind of support were you looking for? If you are comfortable, would you share your posts with me?

\item Why did you decide to choose the Reddit community to seek help?

\item What were others’ reactions/responses to your post?  

\item How did you feel after posting? Did you receive the support you expected?
\end{itemize}
                       
\subsection{Negative experiences}
\begin{itemize}
\item Can you tell me what you think online harm is? Particularly, what online comments or behaviors make you feel uncomfortable or unsafe?

\item Have you ever encountered harmful comments or behaviors in response to your support-seeking posts? Can you tell me more about that?

\item Do you remember how it all started? How did it end? Who was involved? What did you do? What did other people do?

\item What do you usually do if you encounter those comments or behaviors? Why? 
\end{itemize}

\end{document}